\documentclass{article}

\usepackage{arxiv}
\usepackage{float}
\usepackage{cite}
\newcommand{\be}{\begin{eqnarray}}
\newcommand{\ee}{\end{eqnarray}}

\newcommand{\bl}{\color{black}}
\newcommand{\hrd}{\color{black}}
\newcommand{\newc}{\textcolor{black}}
\newcommand{\blue}{\color{black}}

\newcommand{\major}{\color{black}}
\newcommand{\FL}{\textcolor{black}}

\newcommand{\acpt}{\color{black}}
\newcommand{\plag}{\textcolor{black}}

\usepackage{graphicx}
\usepackage{balance}
\usepackage{biolinum}
\usepackage{epstopdf}
\usepackage{moreverb,url}
\usepackage{amsmath}
\usepackage{amsthm}

\usepackage{amssymb}
\usepackage{array}
\usepackage{algorithm}
\usepackage{algorithmicx}
\usepackage{algpseudocode}

\usepackage{subfig}
\usepackage{psfrag}

\algnewcommand\algorithmicinput{\textbf{INPUT: }}
\algnewcommand\algorithmiccons{\textbf{CONSTRAINTS: }}
\algnewcommand\algorithmicoutput{\textbf{OUTPUT: }}
\algnewcommand\Output{\item[\algorithmicoutput]}

\theoremstyle{Definition}
\theoremstyle{Definition} %define the styling for the environment right below it

\newcommand\BibTeX{{\rmfamily B\kern-.05em \textsc{i\kern-.025em b}\kern-.08em
T\kern-.1667em\lower.7ex\hbox{E}\kern-.125emX}}

\usepackage{graphicx}
\graphicspath{ {figure/} }

\usepackage{changes}
\definechangesauthor[name={Yue_Wang}, color=blue]{yw}
\definechangesauthor[name={Fangjian_Li}, color=orange]{fl}

\newcommand{\EditFL}{\textcolor{black}}
\newcommand{\EditFM}{\textcolor{black}}

\newcommand{\EditAA}{\textcolor{black}}

\begin{document}
%\title{Human-Centered Controller Design for Cooperative Adaptive Cruise Control Systems in Mixed Traffic}

\title{A Review of Sensing and Communication, Human Factors, and Controller Aspects for Information-Aware Connected and Automated Vehicles %Design Criterion of Fully Autonomous Vehicles: A Review Study on Data-Driven Connected Autonomous Vehicles \EditFL{Information-Aware?}
}


\author{Ankur Sarker\\
Department of Computer Science\\
University of Virginia\\
\texttt{as4mz@virginia.edu}\\
\And
Haiying Shen\\
Department of Computer Science\\
University of Virginia\\
\texttt{hs6ms@virginia.edu}\\
\And
Mizanur Rahman\\
Glenn Department of Civil Engineering\\
Clemson University\\
\texttt{mdr@clemson.edu}\\
\And
Mashrur Chowdhury\\
Glenn Department of Civil Engineering\\
Clemson University\\
\texttt{mac@clemson.edu}\\
\And
Kakan Dey\\
Department of Civil \& Environmental Engineering\\
West Virginia University\\
\texttt{kakan.dey@mail.wvu.edu}\\
\And
Fangjian~Li\\
Department of Mechanical Engineering\\
Clemson University\\
\texttt{fangjil@clemson.edu}\\
\And
Yue~Wang\\
Department of Mechanical Engineering\\
Clemson University\\
\texttt{yue6@clemson.edu}\\
\And
Husnu S. Narman\\
College of Information Technology and Engineering\\
Marshall University\\
\texttt{narman@marshall.edu}\\
}
\maketitle

\begin{abstract}

Information-aware connected and automated vehicles (CAVs) have drawn great attention in recent years due to its potentially significant positive impacts on roadway safety and operational efficiency. In this paper, we conduct an in-depth review of three basic and key interrelated aspects of a CAV: sensing and communication technologies, human factors, and information-aware controller design. First, different vehicular sensing and communication technologies and their protocol stacks, to provide reliable information to the information-aware CAV controller, are thoroughly discussed. Diverse human factor issues, such as user comfort, preferences, and reliability, to design the CAV systems for mass adaptation are also discussed. Then, different layers of a CAV controller (route planning, driving mode execution, and driving model selection) considering human factors and information through connectivity are reviewed. In addition, critical challenges for the sensing and communication technologies, human factors, and information-aware controller are identified to support the design of a safe and efficient CAV system while considering user acceptance and comfort. Finally, promising future research directions of these three aspects are discussed to overcome existing challenges to realize a safe and operationally efficient CAV.

\end{abstract}


%\begin{IEEEkeywords}
%\sout{Cooperative Adaptive Cruise Control (CACC), Mixed Traffic, String Stability, Human-Centered Controller.}
%\EditFL{ Connected vehicles, automated vehicles, autonomous vehicles, controller, sensing and communication technologies, human factors.}
%\end{IEEEkeywords}

% \fi
%\IEEEpeerreviewmaketitle

%\vspace{0.1in}
\section{Introduction}
%\vspace{0.1in}
 Automated vehicle (AV) field testing began in 1986 in the United States when the Partners for Advanced Transit and Highways (PATH) program at the University of California Berkley developed a platooning application of six AVs in specially guided highway sections \cite{ioannou2013automated}. Since, the most significant AV development was prompted by the Defense Advanced Research Projects Agency (DARPA) Urban Challenge 2007, which accelerated private sector AV research and development. Since then, major automobile companies including internet giant Google have developed the prototypes of AVs that need no special highway infrastructure to operate in mixed traffic scenarios \cite{anderson2014autonomous,mahan15self}. In a research by Bhavsar et al. concluded that although there is a considerable risk of AV sensor failure, future innovations in computation and communication technologies as well as backup sensors can significantly reduce the failure probability of AVs in a mixed traffic stream (which includes AVs and non-AVs)~\cite{bhavsar2017risk}. To facilitate the development of AV technologies, several US states issued special permits to AV technology manufactures conducting pilot testing, most notably in California, automated vehicle laws was issued on February 26, 2018~\cite{CADMV}. This interest in AV technology from both the automotive industry and the public sector will advance the development of fully automated (i.e., autonomous or level 5 automation) vehicle development in the next decade.

The Society of Automotive Engineers (SAE) has a classification scheme for automated vehicles with six levels from no-automation (level 0) to full automation (level 5)~\cite{sae2016taxonomy}. %\cite{smith2013sae}.
Full vehicle automation enables maximum benefit in terms of traffic safety, efficiency, and environmental impacts. According to a recent study \cite{needForSpeed}, AVs require more key insights from different complex, inter-dependent factors of transportation systems (i.e., safety data management and utilization, understanding human driving behaviors, and heterogeneous sensors managements). It is reported that  more than 75\% of US drivers are not comfortable at all to use any kind of AVs \cite{AAA}. Recent tragic crashes of several  automated vehicles highlight the serious life-and-death consequences associated with the uncertainty of traffic environments~\cite{favaro2018autonomous}. Thus, identifying the impacts of three major factors (sensing and communication technologies, human factors, and information-aware controller) on fully automated vehicular systems is necessary.

%However, vehicle-to-X (vehicle-to-infrastructure, vehicle-to-pedestrian, or vehicle-to-vehicle) wireless connectivity along with in-vehicle sensors of AVs can improve the safety and operational efficiency of an autonomous vehicle significantly by having a 360-degree view and by providing support guiding CAVs in case of sensor failures.
{\major

\emph{Firstly}, sensor-based AV with vehicle-to-X (vehicle-to-infrastructure, vehicle-to-pedestrian, and vehicle-to-vehicle) communication can be called as \emph{ Connected and Automated Vehicles} (CAVs). The deployments of vehicular sensing and communication technologies toward CAVs bring significant safety, mobility, and environmental benefit over non-connected automated vehicles~\cite{andersen2013linking}. Sensor and communication technologies enable the  automated vehicles to sense its surrounding environments and communicate with other vehicles or infrastructure such that the AVs would be able to receive/send messages in time and act accordingly. %(e.g., radar, LIDAR, camera, DSRC, and so on)

\emph{Secondly}, it is necessary to study overall human factors to make the riding experience more comfortable and safer from an AV user's perspective~\cite{elbanhawi2015passenger}. In addition, mass adoption of CAVs depends on the user comfort, trust (i.e., accuracy and reliability), and preferences~\cite{teichner1954recent,wilson2010trends}. Any CAV systems must provide a reasonable level of user acceptance. A reasonable level of AV user acceptance depends on the individual's preferences based on their age, gender, cultural, and societal characteristics~\cite{shults2001reviews,rashevsky1966neglected}. A CAV system must ensure acceptable vehicle dynamics (i.e., maximum speed, maximum acceleration/deceleration), headway (i.e., bumper-to-bumper distance between vehicles), gap for changing lane and string stability (i.e., sharp fluctuations of position, speed, and acceleration/deceleration) in different traffic conditions (e.g., congested, free flow) depending on the user preferences.  It is also required to implement driver behavior models (i.e., car-following and lane changing behavior models) for designing a path planning controller to include user preferences as per their expectations \cite{macadam2003understanding}. %

\emph{Thirdly}, the information-aware controller design is critical for safety and efficiency as well as the roadway traffic throughput. Since the V2X communication enables the controller to acquire the information beyond what on-board sensors can detect, it can be expected that the CAV will become safer and more efficient by utilizing the additional information from its surrounding infrastructures and vehicles. Therefore, different control schemes (e.g., model-based predictive controller and learning-based controller) have been developed to utilize the information from external sources. The enhancement brought by the communication technology can be found in every layer of the control structure. Currently, there are many studies seeking to establish the fully automated vehicular systems through analytical and experimental studies. Further advancement of fully automated vehicular systems depends on the current research trends of different aspects of vehicular sensing and communication techniques, human acceptance with their interjections, and information-aware controller designs. }

In this paper, we reviewed existing literature related to CAV systems in terms of the design of sensing and communication technologies human factors and information-aware controller design. {\blue There are several} review studies~\cite{willke2009survey,hsn:karagiannis2011vehicular,hsn:Dey2016control} on different aspects (e.g., communication, controller, and human factors) of future generation Intelligent Transportation Systems (ITS). Our previous survey paper~\cite{hsn:Dey2016control} gives a comprehensive review of the design of Cooperative Adaptive Cruise Control (CACC) systems under the communication effects and consideration of human factors. However, this paper focuses on identifying the effects of sensing and communication technologies, human factors, and controller design criteria for a fully AV design.
%%Identifying the impacts of three major factors (sensing and communication technologies, human factors, and information-aware controller) on fully autonomous vehicular systems is necessary.
%Currently, there are many works seeking to establish the fully autonomous vehicular systems by analytical and experimental studies. Further advancement of fully autonomous vehicular systems depends on the current research trends of different aspects of vehicular sensing and communication techniques, human acceptance with their interjections, and information-aware controller designs. %In this section, we introduce critical future research directions for the advancement of fully autonomous vehicular systems.
%In this paper, we reviewed existing literature related to CAV systems in terms of the design of information-aware controller design, sensing and communication technologies, and human factors. {\blue There are several} review studies~\cite{willke2009survey,hsn:karagiannis2011vehicular} on different aspects (e.g., communication, controller, and human factors) of future generation intelligent transportation systems (ITS). As an example, our previous survey paper~\cite{hsn:Dey2016control} gives a comprehensive review of the design of cooperative ACC (CACC) systems under communication effects and consideration of human factors. However, this paper focuses on identifying the effects of sensing and communication technologies, human factors, and controller design criterion for a fully AV design.

The structure of the paper is as follows: Section \ref{sec:focus} presents the focus of the paper and Section \ref{sec:sensingComminication} presents an overview of vehicular sensing and communication technologies along with the existing research, challenges, and future research directions. Section \ref{sec:human} discusses different human factors to consider designing a CAV and presents related works, challenges, and further research directions on driver behavior modeling for CAV systems. Section \ref{sec:controls} presents a discussion on the information-aware controller designs for CAV systems with a summary on challenges and future research directions. Finally, Section \ref{sec:conclusion} presents concluding remarks of this paper.

\begin{figure*}
\vspace{-0.20in}
	\centering
	\includegraphics[width=0.8\textwidth,height=0.45\textheight]{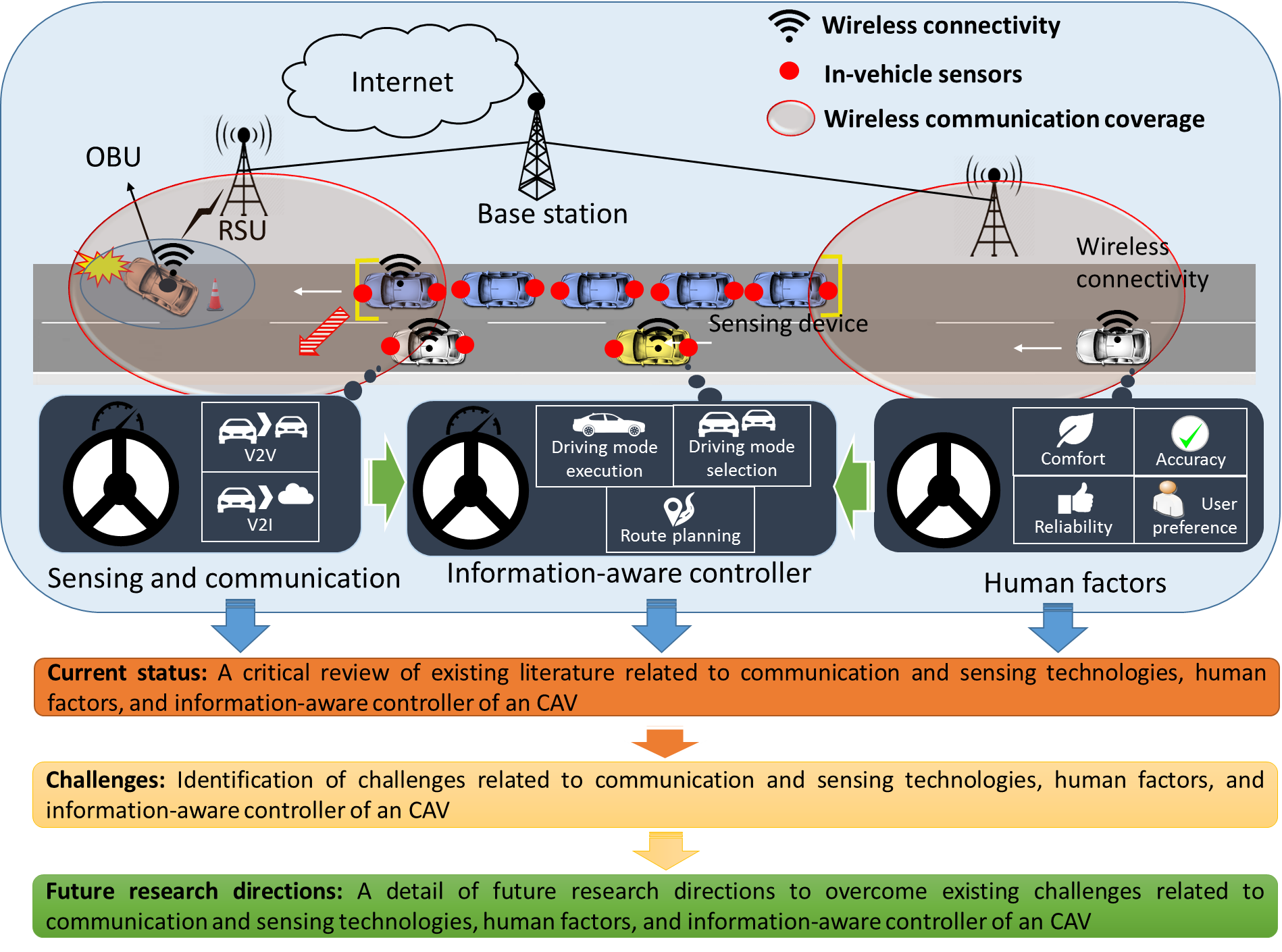}
%	\vspace{-0.1in}
	\caption{ Focus of our paper: Sensing and communication technologies, human factors, and information-aware controller for CAVs.}
	\vspace{-0.25in}
	\label{fig:intro}
\end{figure*}

%\vspace{0.2in}
\section{Focus of the Paper}
%\vspace{0.1in}
\label{sec:focus}
%\vspace{-0.05in}
{\major
The key motivations behind the developments and implementations of automated vehicles are to improve roadway safety and operational efficiency~\cite{V2V2018}. Research and development of CAVs have drawn great attention in recent years from both the industry and academia. However, all of the existing AV developments and real-world implementations (at the time of writing this paper) have been conducted using in-vehicle sensors (e.g., radar, LIDAR, camera, and infrared) assuming that an AV can operate itself by detecting surroundings using in-vehicle sensors. Only relying on in-vehicle sensors will limit the view of an AV and it will not be capable to see what is happening around the corner of a road due to the line of sight constraint, behind an object or further ahead of a road network beyond the sensor range. Vehicle-to-X communication along with the in-vehicle sensors provide a comprehensive 360-degree view to CAVs and it enables AVs' operation beyond the sensor range~\cite{bhavsar2017risk}. Wireless connectivity would also be able to improve the operational efficiency of AVs by providing real-time roadway information, such as traffic signal phasing and timing information, and downstream traffic incident and queue information. In addition, the mass adaptation of AVs largely depends on the consideration of different human factors (e.g., user expectation and ride comfort) in the operational designs of CAVs. The consideration of information obtained through the wireless connectivity and in-vehicle sensors along with the factors related to the user comfort build an information aware AV controller, which provides safer, more efficient, and more comfortable AV operations compared to a conventional AV controller. The driving performance improvements along with the human-factors consideration can together contribute to the well-designed CAV systems, which can contribute to the deployments of CAV systems in different traffic scenarios (e.g., a group of CAVs forming a platoon to increase the safety and efficiency of the road network).

We illustrate a CAV deployment scenario using Fig.~\ref{fig:intro}, where a platoon of CAVs is moving on the right lane and other three CAVs are moving on the left lane. This scenario represents why communication and sensing, human factors, and information-aware controller are the key aspects for safe and efficient operation of an automated vehicle. The platoon of the CAVs is operated by the leader CAV and the platoon is going to change the lane to avoid a roadway traffic incident notified by an roadside unit (RSU). This roadway traffic scenario can be explained by the following actions. The leader vehicle of the CAV platoon receives an incident notification ahead of the roadway incident location from an RSU using V2I communication. The leader vehicle communicates with other CAVs using V2V communication and makes other CAVs of the platoon aware of the hazardous road condition ahead. A reliable and low-latency wireless communication medium is required as a V2V and V2I communication medium to provide the connectivity. Each CAV can also sense its nearby surrounding through onboard sensors. The platoon of CAVs takes appropriate actions such as slowing down and adjust the speed and direction of the entire platoon in real-time.  The leader vehicle of the CAV platoon must consider the user comfort along with the safety of the entire platoon during slowing down and then, lane changing. To ensure the user comfort and safety, all CAVs in the platoon must ensure acceptable vehicle dynamics and string stability of the platoon in the prevailing traffic conditions as per user preferences and comfort. Traffic incident information and user preferences are fed into the CAV controller to calculate the required individual vehicle trajectory that will satisfy user comfort and safety.  As presented in this scenario, sensing and wireless communication technologies, human factors, and information-aware traffic controller are three major factors to be considered for the real-world deployments of CAVs. Thus, the focus of this paper is to present a critical review of the current status, challenges, and future research directions of these three interrelated aspects (sensing and communication technologies, human factors, and information-aware AV controllers) to realize the safe and efficient operations of CAVs.
%}
%}

\vspace{0.05in}
\section{Sensor and Communication Technologies}
%\vspace{0.1in}
\label{sec:sensingComminication}

%It is expected that CAV will provide high standards for safer transportation in different modes of road transposition systems by {\blue intelligently} managing and controlling vehicles and road infrastructure. One of the crucial components which enables this type of managing and controlling mechanism is cooperatively sensing and networking between different components to optimize both common and individual goals.

Using 2004-2008 crash data, a breakdown analysis by the US Department of Transportation (USDOT) states that communication and sensing technologies inside vehicles could help avoid up to 79\% of all traffic accidents  \cite{harding2014vehicle}. Different safety applications (e.g., forward collision warning, emergency electronic brake light, blind spot warning, left turn assist, and lane departure warning) can be offered by sensor and communication technologies for semi AVs or fully autonomous AVs \cite{harding2014vehicle}. %{\blue For example, left turn assist warns the driver of a vehicle, not to turn left in front of another vehicle traveling in the opposite direction.} There are also some other scenarios {\blue where} sensor and communication technologies of vehicles can improve the traffic flow and increase the throughput of the existing road network. As an example, vehicles can be warned by pedestrians crossing a road section.
Additionally, sensor and communication technologies can be deployed to harmonize each vehicle's velocity on the road, control traffic signals, and improve users' riding experience and provide collision warning. %arrange sharing rides,

In this section, we first introduce different sensing technologies, current research trends of these sensing technologies, challenges, and future research directions to increase the safety and traffic efficiency of CAV systems. Then, we present the basics of V2V and V2I communications, existing research trends, challenges, and future research directions on communication technologies. In addition, we {\blue discuss} some promising technologies and their research trends for future generation CAV systems.} %(Section \ref{subsec:future}).
%I have just added the paragraph you need. I still concern about the introduction and our focus section. There are still so many redundancies. You can find the sentence "communication can provide extra information" "human factor needs to be considered for mass adaptation" "controller can improve safety" repeated again and again.

%In addition, for the comments 2" I sometimes miss some more details concerning the content of the cited works, since a survey should not require reading the contents all of the citations. '' I double checked the whole paper. The controller parts give the details. I will modify some if necessary. However, you should double check the following kind of sentences in communication :

%1). Surrounding object detection may be carried out by a different combination of cameras: such as a single camera [26], [27] and multiple cameras [28], [29]. The placement of cameras may be different based upon their purpose.

%2). On the other hand, in contention-based methods, communication channel frequencies, signal power, channel window sizes are adjusted at run time to provide better packet delivery rates [56]â€“[59].

%\subsection{Sensor Technologies}
\noindent \emph{I. Sensor Technologies}\\
\label{subsec:sensor}
%\vspace{0.1in}
%\subsubsection{Current Status}
%\label{subsubsec:future_sensor}
\noindent \emph{A. Current Status:} {\blue Several advanced collision avoidance technologies already available employ different on-board sensor technologies (e.g., RADAR, cameras, and LIDAR) to monitor vehicles' surroundings. These existing ``in-vehicle" technologies are installed inside a vehicle to sense but not communicate with other nearby vehicles~\cite{tuohy2015intra}. RADAR, cameras, and LIDAR installed inside the vehicle are able to collect information directly by sensing the surroundings. As a result, these collision avoidance technologies are able to use surroundings' information to warn the driver about possible hazards so that the driver can take necessary actions to avoid or mitigate the hazards.
}
%next generation iii-i a

%\vspace{0.1in}
\noindent \textbf{Radar.}~Radar emits radio waves to detect the presence of objects by using the time interval between sending radio waves and receiving reflected radio waves. It can also detect the direction of moving objects. There are mainly two types of radar systems: short-range radar (SRR) and long range radar (LRR). SRR can detect objects within 20 meters, uses only a single antenna, and cannot detect angles. It can be used in parking assistance and blind spot warning scenarios. 79 GHz frequency range is used for SRR equipment with a maximum mean power density of -3 dBm/MHz along with a peak limit of 55 dBm~\cite{strohm2005development}. However, LRR can detect any objects within 150 meters with an angular regulation of two degrees. As a result, LRR is able to detect the velocity of objects heading away or toward it. It can be used in forward collision warning and intersection management. Also, the 77 GHz frequency range of LRR equipment allows the combination of -40dBm/MHz transmit power, more than 250 MHz bandwidth, long range operation, and high distance separability at the same time \cite{strohm2005development}. %\EditFL{(Are they the only values? i.e. is 175m, 75GHz possible?)(survey paper)}
For example, radar-based pulse doppler system is able to detect and track objects in front of vehicles~\cite{blanc2004obstacle}. %where the radar system is installed on the lower part of the vehicle.
It can detect objects and their relative speeds within 150 meters based on consecutive echoes of  transmitted radar signals. %This kind of system also detects different vehicles in multiple lanes in real time simultaneously, based on a discrete time signal processing technique, and can detect vehicles in {\blue adverse} weather conditions (fog, rain, etc.) \cite{park2003novel}.

%\vspace{0.1in}
\noindent \textbf{Camera.}~Since cameras are able to detect color and object boundaries, cameras are used to detect the road lanes and read traffic signs. However, cameras can calculate the change rates between objects ahead (e.g., a vehicle gaining speed compared with a slower moving vehicle, or pedestrians, or bicycles). Cameras also delivers unique spatial and color information. Surrounding object detection may be carried out by a different combination of cameras: such as a single camera \cite{tzomakas1998vehicle,van2005vehicle} and multiple cameras \cite{ieng2005new,for2008multi}. The placement of cameras may be different based on their purpose. For example, to detect blind spots, cameras would be mounted nearby the side mirrors and to serve as parking assistance, cameras would be mounted on the back of vehicles. The types of cameras used in vehicles are based on their uses: stereo cameras would be used to obtain wider view \cite{chang2005stereo} while infrared cameras would be used to get a good view at night or during bad weather \cite{niknejad2011vision,kim2012autonomous}.
There are {\blue primarily} three different approaches to detect the moving objects using camera installed in vehicles: background subtraction methods, the feature-based method, and frame inferencing or motion-based methods. A background subtraction method may use a filtering method based on a histogram which collects information from sequences of frames of scatter background \cite{mandellos2011background} or each pixel in the image view to categorize as either noise or a forefront entity's background \cite{magee2004tracking}. In feature-based methods, the nearby objects are discriminated from the background by using their features \cite{papageorgiou2000trainable} and a set of labeled training data are used for feature extraction from the objects \cite{wang2008automatic,papageorgiou2000trainable}. The Haar wavelets technique and support vector machine can be used in these approaches. Similar to background subtraction methods, in frame inferencing or motion-based approach, subsequent frames are compared to extract the background and detect nearby approaching vehicles \cite{mandellos2011background,zhang2010moving,vasu2010effective}.

\noindent \textbf{Light detection and ranging.}~Light detection and ranging (LIDAR) functions similarly to radar systems, as it emits laser signals and uses echoed laser signals to calculate the relative distance of	nearby objects of a vehicle. %We can easily estimate the distances the photos have covered round trip using the speed of light.
LIDAR can measure accurate angles in both horizontal and vertical dimensions and generate three-dimensional data with higher accuracy (within few centimeters error rate) and the generated three-dimensional data are then integrated with two-dimensional GPS data so that vehicles can navigate their surroundings. LIDAR is also utilized for producing high-resolution maps, which are mandatory for AVs to get an overview of their environments. The perception range of LIDAR varies from 10 meters to 200 meters~\cite{thornton2014automated}. A laser scanner is an extension of a laser range finder, which is able to calculate the relative distance of an object using the time-of-flight technique. For example, a two-dimensional LIDAR sensor mounting on a vehicle is used to manage parallel parking. The range estimation capability of the LIDAR sensor is utilized to locate the curb of the road~\cite{thornton2014automated}. %Then, occlusion and location reasoning are used to detect vehicles and their surrounding environments.

%\vspace{0.1in}
\noindent \textbf{Acoustic sensors.}~Mostly used in parking assist, backing, lane maintenance and cruise control, ultrasonic sensors send out high frequency sound waves that measure echoes to determine the distance of an object. For an example, acoustic sensors collect surrounding environment information by received the signals~\cite{sen2011roadsoundsense,mizumachi2014robust}. In \cite{mizumachi2014robust}, an acoustic-based sensing method is presented to extract more robust spatial features from noisy acoustical observations and then, the spatial features are filtered out using sequential state estimation. {\blue The presented system processes acoustic data easily by the mounted microphones outside of the vehicle.} {\blue The spatio-temporal gradient method is used to extract the appropriate features. Then, the spatial features are separated using sequential state estimation.} In another work \cite{sen2011roadsoundsense}, an acoustic sensing hardware prototype is used to estimate congestion on the road {\blue considering acoustic} noises of the surroundings {\blue vehicles}. It samples and processes {\blue acoustic} noise to calculate vehicle speed distribution and {\blue acoustic noise} using differential Doppler shift.

%Usually, vehicle resident sensors would exhibit reduced reliability in certain weather conditions, such as snow, fog, and heavy rain. In addition, camera systems would exhibit reduced performance because of shadows and transitions of light. Majority of existing sensing technologies are susceptible to show poor performance foreign objects, such as snow or dirt. Communication technologies are able to provide safety of CAV systems, increase traffic efficiencies (i.e., flow control), and eventually, complement the sensing technologies of CAV systems.

{\major

%?? ANKUR, THE CHALLENGES MUST BE RELATED TO THE FUTURE RESEARCH DIRECTIONS....WE WILL TALK ??

%\vspace{0.1in}\subsubsection{Challenges}
%\label{subsubsec:challenge_sensor}
\emph{B. Challenges:} %
%
%
%Usually, the in-vehicle sensors are stationary and sensors are typically connected to the electric control unit of vehicle using star-topology network. Also, there is no energy constraint for the sensors having wired connection to the vehicle power system. Data transmissions for in-vehicle sensors require low latency but high reliability to meet the strict requirements of real-time intra-vehicle control system. There are two types of sensors: active and passive. Active sensors are useful to provide real-time detection and they show robustness under bad weather conditions (e.g., rainy and foggy). Active sensors are more exposed to interference issues due to dynamic and noisy environment of road traffic. Noise removal and signal recovery may require complex signal processing techniques. On the other hand, passive vision-based sensors are useful to cover 360 degrees to be equipped on the rear and front side of vehicle. Cameras can track the vehicles moving in a curve or during the lane change more effectively and they can provide more details description of the surroundings. Vision-based sensors are also free from interference problems commonly faced by the active sensors.
%
%
Here are a few challenges for in-vehicle sensor systems:

%\vspace{0.1in}
\noindent \textbf{Fusion of sensors.} The appropriate sensor selection is critical to design a reliable and} safe vehicular system~\cite{tuohy2015intra}. Active sensors (e.g., radar and LIDAR) perform reasonably in different weather/lighting conditions but suffer from interference by other sensors~\cite{thornton2014automated,strohm2005development}. On the other hand, camera-based sensors require higher computational power and they would be sensitive to light conditions~\cite{kim2012autonomous}. Thus, it is important to choose the right sensor groups for AVs based on different objectives including dynamic range, spatial resolution, spectral sensitivity, and computational capabilities.

\noindent \textbf{Security.} Protecting in-vehicles sensors from the cyber-attacks is very crucial for the safety of all nearby vehicles. It may be possible that an attacker can get simply access to the intra-vehicle network to deteriorate the safety and security of all vehicles in the network~\cite{moore2017modeling}.

%\vspace{0.1in}
\noindent \textbf{Line-of-sight problem.} Sensor systems inside vehicle face the line-of-sight problem as they cannot see if there is an obstacle. In addition, interference or obstruction from nearby vehicles is very common in the highly dense traffic scenarios~\cite{thornton2014automated,strohm2005development}.

%\vspace{0.1in}
\noindent \textbf{The complexity of camera sensors.} Camera-based sensors and algorithms should adopt more efficient techniques to handle real-world complex traffic scenarios~\cite{mandellos2011background,zhang2010moving}. Object detection and classification mostly depend on the captured image quality affected by the lighting conditions. Also, vision-based processing algorithms require dedicated and powerful computational resources~\cite{vasu2010effective}.

%\vspace{0.1in}
%\subsubsection{Future Research Directions}
%\label{subsubsec:future_sensor}

%\nointent 
\emph{C. Future Research Directions:} Generally, in-vehicle sensors exhibit poor performance in certain outside weather conditions (e.g., heavy rain, fog, and snow). In addition, the performance of camera systems would be reduced due to shadows and transitions of light.Majority of the existing sensing technologies show poor performance to detect foreign objects (e.g., snow or dirt). %Communication technologies are able to provide safety of CAV systems, increase traffic efficiencies (i.e., flow control), and eventually, complement the sensing technologies of CAV systems.
Based on the above discussion, we can summarize the following future directions to work on in near future for AV sensors.

%\vspace{0.1in}
\noindent \textbf{Fusion of sensors.} Fusion of different sensors have achieved betters results in terms of classification and robustness and they would be appropriate to acquire more detailed and accurate surrounding environments~\cite{thornton2014automated,strohm2005development}. Further extensive research efforts are necessary to design such types of fusion of multiple sensors.

%\vspace{0.1in}
\noindent \textbf{Security.} The security of intra-vehicle wireless sensor network has become a research focus in the recent years and there should be thorough investigations on security aspects of vehicular connected area networks~\cite{moore2017modeling}.

%\vspace{0.1in}
\noindent \textbf{Real-world deployments.}  The majority of the advanced sensors technologies have been tested under different traffic conditions. Further verification is needed to  evaluate these sensor technologies in real traffic scenarios with online traffic data. Also, there should be benchmark studies to compare different sensor technologies~\cite{moore2017modeling}.

%\vspace{0.1in}
\noindent \textbf{Cost reduction.} CAVs require a rich set of different sensors to ensure the safety and comfort of users. This types of sensor combinations should be put into a dedicated hardware unit which could result in more costs and need for more hardware supports. There should be further studies to develop economical and smaller size hardware solutions to make these products affordable.

%\vspace{0.1in}
%\subsection{Communication Technologies}
%\label{subsec:communication}

%\nointent 
\emph{II. Communication Technologies}

%\vspace{0.1in}
%\subsubsection{Current Status}
%\nointent 
\emph{A. Current Status:} Communication technologies are able to provide safety of CAV systems, increase traffic efficiency (i.e., flow control), and eventually, complement the sensing technologies of CAV systems. Studies have shown that cooperation among transportation system components, such as vehicle, transportation infrastructure and driver, in ITS through V2V and V2I communication, improves road safety, traffic flow and air quality, and reduces vehicle energy consumption~\cite{sotelo2012introduction}. There are different components of a fully-integrated vehicular communication system (e.g., a general purpose processor with memory, a radio transmitter and transceiver, several antennas, and a GPS receiver). %It generates the ``Basic Safety Message" (BSM) based on the information gathered from on-board sensors in the vehicle. An integrated system can both send and receive BSMs, and it can process the information of received messages to provide advisories and/or warnings to the driver of the vehicle.

\hrd %Intelligent Transport Systems (ITS) aim to provide a higher standard and safer transportation in different transport models by smartly managing and controlling vehicles and road infrastructure.
%One of the crucial components of ITS which enables this type of managing and controlling mechanism is cooperatively networking between different components to optimize the common and individual goals.
To provide fully-integrated vehicular networking, Federal Communications Commission (FCC) has assigned 75 MHz bandwidth over the 5.85--5.925 GHz for DSRC-based communication~\cite{hsn:FCC-02-302A1:2002}. In DSRC, there are seven channels (172, 174, 176, 178, 180, 182, and 184) with 10MHz bandwidth. Channels 174 and 176 can form channel 175 with 20MHz bandwidth (similarly, channel 181 can be formed by 180 and 182 channels). One of seven channels (channel 178) is called control channel which transmits urgent and management related data, and all other channels are called service channels (SCH)\footnote{While seven channels are allocated in the United States for {\acpt DSRC}, Europe has four channels~\cite{hsn:Campolo2013channel}.}~\cite{hsn:li2010overview}. Table \ref{table:dsrc_requirements} shows the requirements of DSRC system defined by NHTSA. However, the other communication types such as Wi-Fi are not directly suitable for the vehicular environment because of dynamic nature of vehicular traffic and delay sensitive messages that need to be exchanged in a vehicular network ~\cite{hsn:atallah2015vehicular}. Therefore, {\blue physical and medium access control} layers (IEEE 802.11p)~\cite{hsn:802-11p:2010} with additional Open Systems Interconnection layers (IEEE 1609.1, 1609.2, 1609.3, 1609.4, and 1609.11)\footnote{1609.1-1609.4 / architecture, resource manager, security, networking, multi-channel operations, and 1609.11 / over-the-air data exchange protocol~\cite{hsn:li2010overview}}~\cite{hsn:1609} have been designed to address the communication challenges in the vehicular environment. {\blue IEEE 802.11p and IEEE 1609 protocol suites form Wireless Access in Vehicular Environment (WAVE) protocol suite, which determines the architecture and a set of services to enable secure and safe V2V and V2I communications~\cite{hsn:1609}.} In the following subsections, we present the architectural overview of the vehicular {\bl networks}.  Then, we discuss existing V2V and V2I communications.%

%\vspace{-0.1in}
\begin{table}[]
\centering
%\newline
\caption{DSRC-based Communication Requirements.}
\label{table:dsrc_requirements}
\begin{tabular}{|m{0.9cm}|p{2.0cm}|m{1.4cm}|m{2.0cm}|m{1.8cm}|}
\hline
\textbf{Metric} & \textbf{\begin{tabular}[c]{@{}c@{}}Transmission\\range\end{tabular}} & \textbf{\begin{tabular}[c]{@{}c@{}}Elevation\\angle\end{tabular}} & \textbf{\begin{tabular}[c]{@{}c@{}}Transmission\\power\end{tabular}} & \textbf{\begin{tabular}[c]{@{}c@{}}Packet error\\rate\end{tabular}} \\ \hline
Values          & 300m                                                                    & +10 $\sim$ -6                                                    & 15 dBm                                                                 & \textless10\%                                                         \\ \hline
\textbf{Metric} & \textbf{Data rate}                                                      & \textbf{Sensitivity}                                                & \textbf{Bandwidth}                                                     & NA                                                            \\ \hline
Value           & \textgreater6 Mbps                                                      & -92 dBm                                                             & 10 MGHz                                                                & NA                                                                       \\ \hline
\end{tabular}
\vspace{-0.25in}
\end{table}

%\subsubsection{Vehicular Network Architecture}
%\vspace{-0.05in}
In a vehicular network, there might be several vehicle nodes and RSUs which can communicate with each other (as Fig.~\ref{fig:intro} shows). The vehicles are capable of communicating with each other in a short range while they are moving. Here, RSUs are equipped to extend the V2V communication range and provide some other application services (i.e., speed advisory, traffic light managements). As we discussed earlier, the goal of the vehicular communications is to ensure safe and efficient traffic flow. The entire architecture is designed to deliver several kinds of information to drivers, passengers, pedestrians, and vehicles. Currently, any vehicle comes with a rich set of communication units~\cite{allis2015remote} such that {\bl a} vehicle can communicate with other vehicles. For different applications, there would be different application units to store or process the data from communication units and notify {\bl On-Board Unit} (OBU) accordingly. Vehicles can communicate with other moving vehicles which are out of the communication range by using RSUs as relay nodes. Furthermore, RSUs can also be used to connect to the Internet or other gateways (Fig.~\ref{fig:intro}). Thus, any moving vehicle can access the Internet via RSUs.

The followings are the main component of vehicular network:

\vspace{0.05in}
\noindent \textbf{On-board unit.} %\vspace{-0.05in}
An OBU is  basically an IEEE 802.11p enabled communication device which includes a small processor, a memory unit, and a user interface. To communicate with other vehicles or RSUs, an OBU includes an interface which is based on IEEE 802.11x wireless technology. Here, the prior handshaking scheme is excluded from the communication procedures to reduce the communication delay. An OBU may perform several functions (e.g., wireless radio access, ad-hoc routing, geographical routing and reliable message transmission). Inside an OBU, the applications can be mainly divided into two groups: safety-related applications and non-safety related applications. To support the different applications, there is an application unit which is mounted into the OBU; the unit can be a dedicated device for emergency applications or a general purpose internet accessible device. It might be possible that the application unit comes as a part of OBU. Basically, the application unit works as the subordinate of OBU based on application criteria. An OBU {\acpt in} the vehicle is responsible to send Basic Safety Messgae (BSM) containing the vehicle movement information (e.g., current location, speed, acceleration, steering angle, brake status and heading directions) to the nearby vehicles periodically (i.e., at every 0.1 second).
%\vspace{-0.05in}

\vspace{0.05in}
\noindent \textbf{Roadside unit.} %\subsubsection{Road Side Unit}
%\vspace{-0.05in}
An RSU is based on a DSRC device and the communication range of RSU varies from 500m to 1000m~\cite{DSRC}. Since RSUs are static, these would be installed at busy intersections or parking spots where a larger number of vehicles are present and vehicles can have the opportunity to access RSUs. As a backbone of the RSU, there should be some base stations or gateways so that RSUs can be connected to the Internet. Therefore, the RSU can work as a relay node and provide internet connectivity to vehicles. According to~\cite{CAR2CAR}, the main purposes of RSUs include:
\begin{itemize}
\vspace{-0.05in}
\item[] \textbf{\emph{Extending the communication range of V2V network:}} The RSUs can carry and forward messages from one vehicle to another vehicle. Also, RSUs can relay messages to other RSUs and increase the coverage area of the vehicular network.
\item[] \textbf{\emph{Running traffic management applications:}} {\bl The RSUs} may provide special messages to moving vehicles {\blue inside its coverage area about traffic congestion, traffic accidents, hospital zones, etc.}
\item[] \textbf{\emph{Providing internet connectivity:}} Vehicles may connect with RSUs to access the Internet. In this way, RSU may act as a source of information.
    \vspace{-0.05in}
\end{itemize}

In addition, RSUs may run some specific applications, such as eco-driving through signalized corridors, optimum route planner, and network traffic congestion control.
\par
{\bl Due to the advancement of wireless networks, vehicles can connect with cloud infrastructures for accessing cloud services using OBU~\cite{whaiduzzaman2014survey} in order to enhance the network connectivity. The OBU may have the capability to use the Long Term Evolution (LTE) network. {\bl The} vehicles' vendors can make a contract with the nation-wide wireless network providers to install cloud services into their vehicles to enhance vehicle's safety, performance, reliability, etc~\cite{ATnT}. Furthermore, RSUs may connect with cloud infrastructure for getting different cloud services such as Software as a Service (SaaS), Platform as a Service (PaaS), and Infrastructure as a Service (IaaS)~\cite{whaiduzzaman2014survey}. However, the communication procedures between vehicles and cloud infrastructures experience higher transmission delay than other V2V or V2I communications and the communication cost is high.}
%\vspace{-0.2in}

\vspace{0.05in}
\noindent \textbf{Vehicular networks.} In an automated vehicular system's communication network, there are two main issues to handle:
\begin{itemize}
%\vspace{-0.1in}
\item Message dissemination among vehicles inside or outside of the communication range,
\item Better communication schemes for V2X (V2V and V2I).
%\vspace{-0.1in}
\end{itemize}
In an automated vehicular system, vehicular communication can be divided into two main categories which are discussed in the following subsections.

\vspace{0.05in}
\noindent \textbf{\emph{Vehicle-to-vehicle communication.}}
%\vspace{-0.05in}
%Beacon control
In a highly dynamic environment, each vehicle in a single lane, follows a leader vehicle in a Cooperative Adaptive Cruise Control (CACC) environment. Thus, to facilitate the stability of the traffic system, each vehicle needs to periodically transmit its current position, velocity, and other information to its neighbor vehicles, which is called beacon message. To facilitate the continuous transmission of beacon messaging, several methods have been presented which can be further categorized into two types: contention-free and contention-based. In the contention-free beacon message dissemination, vehicles are arranged in several groups and the communication slots are divided into different time slots, called Time Division Multiple Access (TDMA)~\cite{ahizoune2010contention,almalag2012tdma}. On the other hand, in contention-based methods, communication channel frequencies, signal power, and channel window sizes are adjusted at run time to provide better packet delivery rates~\cite{rawat2011enhancing,wang2012blind,park2013collision,park2012application}.

%congestion control
There are several efforts~\cite{sepulcre2011congestion,stanica2011local,bansal2013limeric,sarker2016decentralized} that have been made to reduce the channel congestion of V2V communication network. Proactive and reactive controllers have been investigated for beacon congestion control system using distributed manner~\cite{sepulcre2011congestion}. Here, the proactive controller estimates the desired transmission parameters using current neighbor vehicles and the reactive controller is the feedback-based controller to provide the  transmission robustness. In another work~\cite{bansal2013limeric}, a linear message congestion control technique has been presented where the packet transmission rate is controlled by using feedback messages from neighbor vehicles. However, this work is limited to {\bl the single-hop} scenario {\blue (i.e., no intermediate message relay nodes)}, {\bl which means it does not work for ad-hoc networks}. This research by Stanica et al. {\bl \cite{stanica2011local}} presents the effects of contention window on inter-vehicle communication where the authors discuss several approaches to adjust the minimum congestion window based on roadway traffic density such that the performance of the IEEE 802.11p protocol would be improved.
%platoon features

%\EditFL{Why emphasize Cruise control, platoon? how about other functions?}
CACC or platoon systems highly rely on V2V communications to maintain the stability of the system. There are other sets of research which use cruise control features to avoid the collision of the communication channel. For example, in the work~\cite{segata2014towards}, the leader vehicle first transmits the message to avoid the contention with other vehicles. Then, all other vehicles transmit messages based on TDMA approach where only leader vehicle is capable of communicating with other vehicles. Other vehicles can only communicate with its nearby neighbor vehicles. Amoozadeh et al. presented a platoon based communication protocol where vehicles are {\bl connected} through an ad-hoc network~\cite{amoozadeh2015platoon}. In {\bl that} approach, vehicles inside the platoon can dynamically perform three types of maneuver: joining to platoon, leaving from platoon and lane changing for the entire platoon. For platoon maintenance, this system uses vehicles' control logic considering intra-vehicle distance and speed, and acceleration of vehicles. Also, a separate beacon message is designed and single hop message transmission is allowed to alleviate the communication cost of vehicles. Another work~\cite{fernandes2012platooning} uses the vehicles' relative position information with respect to {\bl the} leader vehicle position to decide which {\bl vehicles} can transmit {\bl messages} at a particular time slot. In the work~\cite{jonsson2013increased}, the entire platoon is distributed into different regions based on the communication range of vehicles. One master vehicle is selected  in each region to coordinate the message disseminations for collision avoidance and {\bl enlarging} the platoon length as well. However, to ensure the connectivity of the whole platoon, packet retransmission is supported in the transmission layer.
%change

In another research of platoon based vehicle ad-hoc network~\cite{shao2015performance}, connectivity probability is studied based on different parameter settings (e.g., roadway traffic density, the communication range of vehicles and RSU, inter-RSU distance, etc) to design a connectivity-aware MAC protocol. This protocol works in a multi-channel reservation based system, and it considers both roadway traffic density and connectivity state to adjust the transmission rate of the beacon message. To further ensure the safety of vehicles in {\bl a} platoon, a multi-priority Markov model is used to analyze the performance of underlying network connectivity and packets belong to intra-platoon transmission are assigned a higher priority if necessary. However, the work considers, {\bl the} platoon formation is static and vehicles cannot leave or change the lane.

Reducing channel {\bl congestions} in V2V communication is a challenging task {\blue where roadway traffic density} is high and vehicles are always moving. It is an open research problem to design effective periodic {\bl messages} or beacon transmission {\bl schemes} for CAVs as it needs consideration of such traffic conditions, which is always changing and the underlying communication channel is not stable or reliable.
%\vspace{-0.10in}

\noindent\textbf{\emph{Vehicle-to-infrastructure communication.}} %\subsubsection{Vehicle-to-Infrastructure Communication}
%\vspace{-0.05in}
V2I communication {\bl in CAVs} {\blue takes} place to {\bl favor} each vehicle to {\bl establish} a stable communication network. Mainly, V2I communication involves vehicles and RSUs which communicate with each other. Usually, the transmission delay of V2V communication is much shorter than the transmission delay of V2I communication but V2V communication is not reliable and stable as vehicles are always moving {\bl and the inter-vehicle distances are changing as well. Also, the transmission of messages from a platoon system may experience signal interference with the messages from another platoon system if two platoon systems are close to each other or crossing each other. However, two platoon systems may share information about hazardous road conditions or upcoming traffic congestions, which is only possible if these platoon systems are nearby to each other.}
To mitigate these issues, there would be several DSRC-based RSUs installed in each traffic intersections so that vehicle can communicate with each other via RSUs if they are out of each other's communication range.
%RSU as rely node
In the work~\cite{sou2011enhancing}, the RSUs are used as forwarders to relay safety message between different group of vehicles. Abdrabou et al.~\cite{abdrabou2011probabilistic} estimated the minimum number of RSUs for a particular road by considering the delay of multi-hops packet delivery for V2I communication. Another similar work by Zhang et al.~\cite{zhang2012multi} presents the performance of uplink and downlink connectivity between vehicle and RSUs in the ad-hoc mode. It also investigates the features of inter-RSUs distance, vehicle density, radio coverage and the maximum number of hops for connectivity between vehicles and RSUs.
%issue of RSU

Typically RSUs are equipped with the 802.11p-based DSRC devices where IEEE 802.11p uses carrier sense multiple access with the collision avoidance (CSMA/CA) mechanism. Several research try to address the issues of RSUs (i.e., high channel congestions) when vehicle density is too high. In a dense roadway traffic scenario, the DSRC-based devices would show a poor performance with higher packet loss and average delay~\cite{han2012analytical}. In the work~\cite{han2012analytical}, a V2V message forwarding scheme is designed to extend the coverage range of RSUs, improve the link quality and maintain high throughput. It chooses the intermediate-destination vehicle node to forward the message based on platoon's velocity. Jia et al.~\cite{jia2014improving} analyzed the performance of drive-through internet to present a platoon-based cooperative re-transmission technique with the consideration of roadway traffic mobility and wireless communication together. In their approach, each vehicle helps to forward the data on behalf of its neighbors in the case of the transmission failure and the cooperative re-transmission behavior is modeled using a 4-D Markov chain formulation. Bi et al.~\cite{bi2013medium} presented an IEEE 802.11e based MAC layer protocol for V2I communication to guarantee a minimum delay for emergency messages while maintaining high QoS performance for other messages. Basically, in their work, there are several vehicles and RSUs are distributed randomly and they consider the busy tone signaling in MAC protocol to consider high priority for emergency applications. The communication channels are divided into two groups: busy tone channel and data channel. On the busy tone channel, the algorithm transmits channel jamming signals, which is called ``busy tone". This method can be applied to the automated vehicular system where vehicles can group together and one leader vehicle contacts with RSUs for {\blue receiving} services.

%\EditFL{Why emphasize Cruise control, platoon? how about other functions?}
In \cite{zhao2008extending}, platoon features are used to meet the Quality of Service (QoS) of vehicular applications. Based on the proximity of nearby vehicles, a group of vehicles forms a platoon system where different channels are used for inter-platoon and intra-platoon communications. A hierarchical optimization model is designed to maximize the utility of an individual vehicle inside a platoon and to minimize the cost of reserving a stand-by channel based on data transmission and collision threshold. The connectivity probability  (i.e., probability of active connection) of V2X is thoroughly investigated in the work~\cite{shao2014analysis} where the vehicles are considered as Poisson distributed with different traffic densities. The work also includes the relationships between the connectivity probability and other parameters, (e.g, vehicle density, the transmission ranges of different elements in the network) to ensure the connectivity.

A group of the existing work~\cite{naik2017coexistence,dey2016vehicle} also consider the hybrid wireless V2X communications consist of DSRC, LTE, and Wi-Fi technologies. In the work~\cite{naik2017coexistence}, the co-existence of 802.11p-based DSRC and 802.11ac Wi-Fi technologies is considered by sharing the 5.9 GHz band so that vehicles can be connected with the commodity Wi-Fi devices. Dey et al.~\cite{dey2016vehicle} considered the hybrid network topology based on DSRC, LTE, and Wi-Fi so that the overall network coverage can be extended beyond the communication range of DSRC.

Furthermore, there are existing work~\cite{lee2014vehicular,yu2013toward,gerla2014internet} {\blue that} consider vehicle-to-cloud communication networks. For example, the work~\cite{lee2014vehicular} considers vehicle cloud models and data routing and dissemination techniques for the vehicular ad-hoc network. The vehicle cloud model is dynamic, created by cooperatively sharing available resources from vehicles and RSUs. Besides, it envisions vehicular cloud networking and encourages collaborations among cloud members (vehicles/RSUs) to provide advanced vehicular services. Due to its resource sharing properties, this work can be incorporated in automated vehicular systems where vehicles cooperate with each other. Another work~\cite{yu2013toward} also discusses the opportunities to establish the local vehicle cloud, road side vehicle cloud, and remote vehicle cloud based on V2X networks. Here, vehicles share resources with each other to create a local vehicle cloud, a road side vehicle cloud is created based on available resources of RSUs and a remote vehicle cloud resides on some remote servers or data centers.

%From above discussion, we can summarize that V2I communications and vehicles' mobility models are closely related to each other. {\blue Several previous studies have been proposed to tackle the poor performances of V2V communications caused by vehicles' mobility models using RSUs.}
 %%RSU
\vspace{0.05in}
\noindent \textbf{Next generation communication technologies.} \label{subsec:future} %{\acpt ????????????????ANKUR, Dr Chowdhury TOLD US TO MOVE THIS SECTION TO CURRENT STATUS AS I SAID BEFORE???????????????.....PLEASE DO IT?????????????}}
In this subsection, we present some promising future generation communication technologies for CAVs:
%\vspace{0.1in}

\noindent \textbf{\emph{Visible light communication.}} Light-Fidelity (Li-Fi) is a wireless communication technology and it uses the band of visible light for data transmission. Li-Fi is faster than other wireless communication, it is useful in secure communications as light cannot penetrate strong objects, and it is inexpensive due to the cost of LED lights. The data transmission is carried out by LEDs' flickering states. Due to vision persistence of human eyes, Li-Fi data transmission is not undetectable for human. Different strings of 0's and 1's can be decoded to retrieve the transmitted information.
A LED can act as a sender and a silicon photo diode can act as a receiver. Different data modulation techniques are used for Li-Fi devices to achieve data transmission range up to 40 Mbits/s \cite{lourencco2012visible}. Li-Fi typically uses visible light between the wavelength 780 nm and 375 nm.

%The cost of Li-Fi is less than other communication techniques as LEDs have been commonly used in automotive lighting \cite{scopigno2015potential}. In contrast to the typical V2V communication strategy, the work~\cite{abualhoul2013platooning} studies the feasibility of visible light communication strategy based on Bit Error Rate (BER) showed poor performance in the presence of longer inter-vehicle distance, background noise, incidence angle, and receiver's electrical bandwidth. In the method, the vehicle's rare light modules are used as a communication module and the authors evaluated the proposed method by using communication channel's DC gain model, noise model with vehicles' trajectory control theory. In \cite{shin2011vlc}, bi-directional Li-Fi transceiver is implemented using edge emitted laser diode and silicon photo diode for short range data communication. Based on the implementation, data transfer can be operated in full duplex mode at 120 Mbits/s. Li-Fi transceiver is also proposed for V2V communication based on 802.11 MAC protocol \cite{tomavs2014simulating}.

%\vspace{0.1in}
%\subsubsection
%\noindent
%for cavs??
\noindent \textbf{\emph{LTE Advanced Pro.}} LTE Advanced Pro (LTE-A) is the evolutionary path from LTE Release 14. LTE-A provides access to a wide range of packet-based telecommunication services. The goal is to reduce the idle time from 100 ms (currently in LTE) to less than 50 ms. Similarly, the transition time from the dormant connected node to the active connected node should be reduced from 50 ms to less than 10 ms. Radio communications have already been shown to improve road safety and traffic flow efficiency, and radio communications are important for the deployment of CAVs. To support different CAV  applications by satisfying communication reliability requirements (e.g., communication latency  requirement and reduction of data loss rate) in a real-world environment, the 3rd Generation Partnership Project is developing different solutions for different V2X communication scenarios, including V2V, V2I, vehicle-to-pedestrian (V2P), and vehicle-to-network (V2N) based on {\acpt the} LTE network. LTE V2X  specifies only the lower layer protocol, and it reuses higher layer protocol and services of LTE network.

%, and hence specify only the lower layers. %An LTE solution will be able to utilize the existence of an already deployed network infrastructure to support many of the use cases and provide an increased level of security for distributed systems.
The report~\cite{3gpplte} is published to include a wide range of groups characterizing the different service requirements: authentication, capacity, service charging, and so on.

{\major
%\vspace{0.1in}
\noindent \textbf{\emph{Fifth Generation Wireless Communications.}} %{\acpt ????? NEED TO REVISE THIS SECTION --- I HAVE ALREADY REVISED ---I WILL ATTACH A DOCUMENT ABOUT 5G---DR c SUGGESTED TO ADD DESCRIPTION BASED ON GIVEN DOCUMENTS??? ANKUR PLEASE ADDRESS IT????}
The ``5G/IMT-2020 Standing Committee" reports the standards and projects of plausible relevance to fifth generation wireless communication (5G) technologies~\cite{IEEE8025G}. The IEEE 802.11 working group develops the standards for IEEE 802.11ad to be used in the 5G standards with the follow-up protocol P802.11ay, which also supports high individual throughput in the millimeter-wave bands as the part of 5G technologies. The IEEE 802.16 WG also develops specific standards and requirements for wireless MANs. Following the IMT-2000 (3G) and IMT-Advanced (4G), 3GPP has been working on the developments of IMT-2020 (5G) since September 2015. 3GPP Release 15 defines the 5G standards~\cite{IMT2018}, where total carrier Bandwidth is 100MHz for a single gigabit backhaul and total carrier Bandwidth is 500MHz for a multi-gigabit backhaul, the data rate is about 10 Gbps, estimated latency is less than 1 ms round trip time, and the frequency spectrum is around 450 MHz to 6 GHz. Also, 5G should be backward compatible with current LTE and its control plane is same as LTE. 5G New Radio is expected to expand and support various application scenarios such as enhanced Mobile Broadband (eMBB), Ultra Reliable Low Latency Communication (URLLC), and massive Machine Type Communication (mMTC). eMBB is designed for the mobile broadband services, which requires high data rate along with seamless data access in different environments (e.g., indoor and outdoor facilities). In addition, URLLC is defined for the applications in high mobility vehicular communication scenarios, which require strict communication latency and reliability requirements to enable CAV communication network.  %In a high-mobility scenario where nodes are moving with velocity up to 500km/h, a higher density of reference symbols would be configured for robustness against a faster time-varying channel.  A smaller periodicity of the slot is preferred for low latency and a self-contained slot structure is suggested due to its fast Acknowledgement/Non-Acknowledgement for data transmissions. Moreover, 5G New Radio hopes to support URLLC services and eMBB services simultaneously on one carrier. URLLC is implemented with a higher priority with guaranteed time-frequency resource by puncturing the eMBB services. To summarize, flexible design and configuration of the transmission modes and parameters are the core idea of 5G New Radio.

}

{\major

%?? ANKUR, THE CHALLENGES MUST BE RELATED TO THE FUTURE RESEARCH DIRECTIONS....WE WILL TALK ??
%\vspace{0.1in}
%\subsubsection{Challenges}
%\label{subsubsec:challenge_sensor}
%\nointent 
\emph{B. Challenges:} %
%Overall, vehicular communication presents unique characteristics of high mobility of communication nodes and this type of communications fall into two categories: V2V and V2I. DSRC-based V2V communication is able to ensure the safety application requirements when the traffic flow is normal. Cellular-based V2V communication would be useful in the scenarios when the congestion is higher. Cellular-based V2V communication can provide higher network capacity to support of high bandwidth demand and data-hungry applications. It can also have wider coverage range to reduce the frequency of horizontal handovers since the vehicle-to-base station communicaiton time is relatively longer than DSRC-based V2V.
%On the other hand, DSRC-based V2I communication is enabled to increase the overall coverage area of the communication network and it helps to enable outside world with vehicles.
Here are some few challenges of the existing vehicular communications:

%\vspace{0.1in}
\noindent \textbf{Dynamic vehicles movements.} The network topology of vehicular communication changes very frequently due to high vehicle mobility and different trajectories of different vehicles. It can result in the frequent data flow disconnections. To solve this problem, a vehicle can utilize a multi-hop communication to transmit messages to another vehicle, which is outside the DSRC coverage area of that vehicle~\cite{naik2017coexistence,dey2016vehicle}. %(by allowing vehicles to relay packets to/from the gateway), the
However, the existence of a data path between two vehicles is not guaranteed, especially,in low number of CAVs in the traffic stream scenarios where two nearby vehicles would be out of the communication range. The packet routing is also a challenging task due to the highly dynamic network topology.
%

%\vspace{0.1in}
\noindent \textbf{Channel congestion.} Cross-channel interference in DSRC-based V2V communication introduces packet drop rates when two adjacent channels are operated simultaneously~\cite{sepulcre2011congestion}. In a higher {\acpt roadway} traffic density situation, the {\acpt communication} channel congestion intensity among nearby vehicles increases remarkably and it results in higher transmission collision rate and a larger channel access delay~\cite{stanica2011local}.} In addition, due to the lack of handshaking and acknowledgement procedures, the performance of DSRC communication is degraded  significantly while delivering broadcast frames in a higher traffic density and it results in an unreliable broadcast service due to the hidden terminal problem where two or more vehicles (i.e., outside of each others' communication ranges) send messages to a common vehicle at the same time~\cite{bansal2013limeric}.

%\vspace{0.1in}
\noindent \textbf{Latency of cellular V2V.} All data in cellular networks need to go through by the base station due to the centralized control, which restricts its usages for the safety applications that have strict delay requirements~\cite{dey2016vehicle}. The downlink channel becomes a bottleneck even when the number of vehicles in a region is small. In some scenarios, the base station would broadcast a safety message to all the vehicles in a region even though it may be irrelevant to some vehicles. Therefore, vehicles need to perform unnecessary processing to understand the significance of many unwanted messages. One solution would be to use the multicast service and sending messages to a specific group of vehicles. However, the overall cost of this solution would be high considering latency and control signaling overhead due to the the group creation and maintenance. In addition, the uplink channel is congested in a high density of CAVs in the road network scenario as uplink transmissions are always enabled using the unicast mode.

%\vspace{0.1in}
%\subsubsection{Future Research Directions}
%\label{subsubsec:future_sensor}

%\nointent 
\emph{C. Future Research Directions:} %
The advancement of communication architectures makes it possible to support safety applications for existing transportation systems. However, current vehicular communication is still not ready to ensure stability and safety of a well maintained fully automated vehicular system. The uncertainty of traffic conditions would hinder reliable and continuous connectivity of V2X communications. Besides, network security is crucial for the safety of traffic environment. Several future research directions for fully  automated vehicular systems are introduced as follows.

\noindent \textbf{Heterogeneous vehicular communication.} To provide user comfort and ensure traffic safety, both V2V and V2I communications are required. IEEE 802.11p-based DSRC communication is the backbone of the existing vehicular communication and it works well for static fully  automated vehicular systems. However, the limited coverage of RSU units brings the necessity of heterogeneous vehicular communications. The research underway into heterogeneous network standards may form the basis for future ITS applications. The heterogeneous communication protocols and schemes should be robust and scalable to deal with the uncertainty of traffic conditions~\cite{naik2017coexistence,dey2016vehicle}. The resource scheduling unit should consider application criteria, synchronization, and a context-aware mechanism to provide more dependability of heterogeneous communication links for fully automated vehicular systems. Current transmission control protocols should be customized to be suitable for robust and highly scalable data dissemination scenarios.

\noindent \textbf{Signal interference avoidance.} In congested roadway situations, the signal interference avoidance for BSMs is a critical research direction to increase the reliability of automated vehicular systems. The optimized V2V channel access mechanisms should be considered to satisfy stringent latency requirements. There are already a lot of theoretical work that consider the efficient channel allocation mechanisms based on message sending rate, duration of contention windows, channel switching delay, successful message reception rate, channel busy ratio, throughput, packet error rate, and so on~\cite{sepulcre2011congestion,stanica2011local,bansal2013limeric}. However, the channel allocation mechanisms should be evaluated and verified in different real world roadway traffic scenarios considering uncertain nature of vehicular communication in fully  automated vehicular systems.%

\noindent \textbf{Security and privacy in vehicular communication.} The heterogeneous connectivity of vehicles inside an automated system demands strong security mechanisms to prevent unwanted access to vehicles' control. Messaging between vehicular systems could be received, tempered or altered by malicious nodes. The centralized certification revocation in automated systems would cause longer delay and interrupt the sporadic connections between leader and follower vehicles. Further research should consider developing security architecture to support heterogeneous network architecture considering reliability and QoS of automated vehicle systems. The combination of physical and application layer security mechanisms might be useful to reduce the overhead to impose the security of automated vehicle systems. Besides, the architecture should support the anonymity of vehicles outside of automated vehicular systems such that personal information can not be identified. The data encryption mechanisms considering computing resource, time, and network architecture for AVs would be a research direction to follow.\looseness=-1

In next section, we discuss the human factors involved in the design process of CAV systems.

%\vspace{0.05in}
\section{Human Factors}
\vspace{0.1in}
\label{sec:human}
\begin{figure*}
\vspace{-0.20in}
	\centering
	\includegraphics[width=1.0\textwidth,height=0.21\textheight]{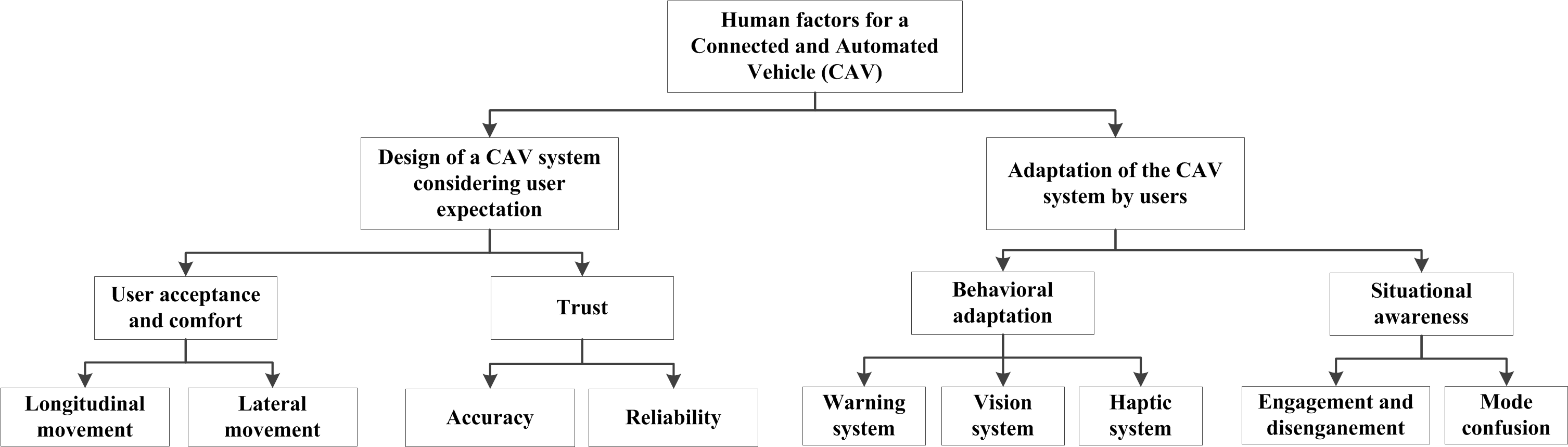}
%	\vspace{-0.1in}
	\caption{Human factors for connected and automated vehicles.}
	\vspace{-0.15in}
	\label{fig:humanfactors}
\end{figure*}

Every year, thousands of people die from traffic {\blue incidents} and the major contributing factor identified in numerous studies is driver errors. To encounter and support driver mistakes/limitations, such as long reaction time to unexpected/expected roadway events \cite{teichner1954recent}, distracted driving \cite{wilson2010trends}, and driving under influence \cite{shults2001reviews}, modern vehicles are equipped with advanced driver assistance systems, such as lane keeping assistance, blind spot warning and forward collision warning. However, new features such as entertainment system, hand held devices (e.g., smart phone and tablet), navigation system, create new forms of distracted driving and been reported as the primary cause of thousands of crashes. While intervention of more technology is future vehicle models are unavoidable, the promise of fully automated vehicle is that drivers {\blue do not} need to control the vehicle and most of human mistakes and driver distractions could be eliminated by automated vehicle control system. The feedback between vehicle's operational environment and associated driver behavior follow a complex pattern \cite{rashevsky1966neglected}.  While understanding driver behavior and modeling is recognized as a complex issue in traditional vehicles, it is also a key challenge for automated vehicle design \cite{macadam2003understanding}. Thus, the interactions between human and vehicle have been a core research focus of the automobile industry and academia. It is imperative to understand human behavior in designing key features of CAV systems and reliable human-machine interfaces.
\vspace{-0.15in}
\subsection{Current Status}
The mass adaptation of CAVs largely depends on the how the automated system can be designed based on the human factors, such as user expectation, ride comfort, and user trust on the automated system \cite{wang2015human,blanco2015human}. The report entitled \emph{Human Factors' Aspects in Automated and Semi-Automatic Transport Systems: State of the Art} identified major human factor issues, which are: acceptance and comfort, situational awareness, loss of skill, behavioral adaptation and risk compensation, workload, level of automation, and normal transitions, responses to system failures, usability, and guidelines \cite{martens2008human}. %\EditFL{Is this for the fully autonomous car?}.
Human factors consideration for CAV can be broadly categorized into two groups: i) design of a CAV system considering user expectation, and ii) adaptation of the CAV system by users (as shown in Figure \ref{fig:humanfactors}). The considerations for human drivers' expectation in designing the CAV system can be further classified into two sub-groups: a) user comfort and acceptance and b) user trust. On the other hand, users of the CAV system need to adapt the engineered and designed CAV system, which includes a) behavioral adaptation and b) situational awareness. An overview of human factors' consideration in CAV design is illustrated in Figure \ref{fig:humanfactors} and discussed in the following subsections.

\vspace{0.1in}\subsubsection{Design of a CAV System Considering User Expectation}
Understanding how human will interact with the automated system is an important research focus of a CAV system. Satisfying user requirements in terms of comfort, workload, perception-reaction, and maintaining safety are extremely critical in designing the safe and reliable CAV. This subsection provides a detailed review on the user comfort and acceptance, and trust of the CAV systems.

\vspace{0.05in}\noindent \textbf{User comfort and acceptance.} User comfort experienced in a CAV system will be an important factor in terms of user acceptance.  A CAV controller operates the longitudinal (i.e., car-following model) and lateral (i.e., lane-changing mode) movement of a vehicle, and must replicate a human driving experience in each CAV where expectations of a user for comfortable driving experience is not violated. Longitudinal driving behavior models and lateral driving behavior models capture human driving behaviors in different driving environments. Car-following models represent driver's reactions to the surrounding environment in the car-following mode. Car-following model must ensure acceptable vehicle dynamics and string stability while CAVs are in a traffic stream to improve user acceptance, comfort, and safety. Car-following models capture how a subject vehicle follows the preceding vehicle by maintaining longitudinal position and minimum gap, and driver's reaction for the longitudinal movement of a vehicle. The acceleration/deceleration behavior of a CAV's car-following model must maintain comfortable acceleration/deceleration in different driving conditions. Thus, the car-following behavior for the longitudinal movement needs to be examined in three different acceleration/deceleration scenarios for the CAV controller design: i) no acceleration/deceleration (uniform speed); ii) a constant deceleration; and iii) a constant acceleration \cite{Rahman2017Evaluation}.

Different existing car-following models, which capture human driving behaviors, can be examined for a CAV controller design for longitudinal movement control to achieve user comfort and acceptance of a CAV system. The general form of car-following models assumes that each driver reacts to a stimulus, which leads to an actuation of the acceleration/deceleration \cite{brackstone1999car}. Many car-following models, such as the Gazis-Herman-Rothery (GHR) model, the Collision Avoidance (CA) model, the Helly model, the Fuzzy Logic based model, the Optimal Velocity (OV) model, and the Meta model, have been developed since the 1950s \cite{brackstone1999car}. Despite the substantial research related to GHR model dating to 1958, the many contradictory findings regarding the correct parameter selection have resulted in substantially less GHR follow-up work. It {\blue calculates} a {\blue speed with a} safe space-headway to avoid collision with the preceding vehicle. The Helly model, also known as the linear model, was developed based on the GHR model \cite{helly1900simulation}. Although the Helly model fits well with observed data, the calibration of model parameters is the main difficulty because of a large set of parameters \cite{panwai2005development}. In the 1990s, the fuzzy logic theory was introduced to model car-following behaviors to better consider the fuzziness in drivers' decision-making processes \cite{kikuchi1992car}.  The fuzzy rules capture the reactions of a driver to the actions of other drivers based on a set of rules, which are developed through driving experience. {\blue Thus, fuzzy logic based car-following models could capture the CAV users' characteristics for the car-following mode.} In the OV model, the acceleration/deceleration of a subject vehicle is determined by a function, which is consists of the optimal speed of a vehicle and driver sensitivity \cite{gong2008asymmetric}. The uniqueness of the OV model is that it can capture the car-following behavior of a vehicle at different levels of traffic congestion on a roadway. Wiedemann presented a psychophysical car-following model based on a perceptual threshold of the driver to model different types of driving regime (e.g., a free driving regime where time headway is larger than a predetermined threshold and a emergency driving regime where time headway is lower than a predetermined threshold to avoid a possible collision with the lead vehicle) \cite{wiedemann1974simulation}. In this model, the perceptual threshold of the CAV users for each driving condition depends on the gap and the relative speed between the subject and the preceding vehicles, and assumes that the CAV will react when they reach these thresholds. Gipps presented a multi-regime car-following model for congested and free-flow traffic conditions \cite{gipps1981behavioural}. The maximum acceleration of a vehicle for these traffic conditions being determined according to the following two conditions: i) the drivers desired speed, which is the speed limit of a roadway; and ii) the minimum space-headway, which is required to avoid collisions. The second constraint excludes the chance of accident occurrence of a CAV when compared to actual car-following movement in traffic. Yang and Koutsopoulos presented a multi-regime model, which also precludes incident-inducing car-following mode \cite{yang1996microscopic}. The Intelligent Driver Model (IDM) is another multi-regime model, which captures the dynamics of different traffic congestion level more realistically than any other models \cite{treiber2000congested}. According to this model, the acceleration of a subject CAV can be designed based on the subject%??can you say like the depends on a specific/certian vehicle-better to generalize??
CAV's speed, the ratio of current gap and the desired gap between subject and the preceding CAVs, and the relative speed between the subject and preceding CAVs. This car-following model was used as in car-following mode controller for the vehicle automation for CACC system design. For example, Milanes and Shladover evaluated three controllers using field data in 2014: i) ACC system of Infiniti M56s test vehicle; ii) CACC control systems, which is developed by Milanes, et al.~\cite{milanes2014cooperative}; and iii) CACC control systems using the IDM \cite{milanes2014modeling}. They have used their production vehicles for the field experiments. The actual responses of the vehicles and users were measured, and it was found that the IDM model demonstrated comfortable car-following behavior than the other controllers. However, the IDM model shows slower response and large space-headway between CACC vehicles. {\blue T}he critical factors considered in existing car-following models are identical: the subject vehicle's own speed difference, the distance between the subject vehicle and the one it follows, and driver's reaction time \cite{Rahman2017Evaluation}. However, evaluation of vehicle dynamics and string stability of a car-following model is critical to assess the user comfort and acceptance. The following paragraph introduces the existing lane-changing methods and discusses the limitations and challenges of these models in designing the CAV systems.

Merging to and diverging from a lane is related to lane changing behaviors of a CAV user. For example, as a CACC system allows a minimum space-headway between vehicles; it becomes very challenging for a vehicle to join an existing platoon at any location other than at the beginning or end of the platoon \cite{jones2013cooperative}. Similarly, a vehicle attempting to leave a platoon will likely need to adjust its desired speed and gap depending on the user preferences and comfort to prepare to move to an adjacent lane. A CAV lane-changing model must replicate user preferences considering the modeling of user characteristics, such as the variability of CAV user behaviors across different user types (i.e., younger/older and aggressive/non-aggressive drivers), users gap acceptance behavior, and gap availability in the target lane \cite{do2017human,rahman2013review}. In terms of vehicle's lateral movement control, numerous researchers presented different models in terms of modeling lane-changing behavior. Chee and Tomizuka compared linear quadratic (LQ) controller, frequency shaped linear quadratic optimal control (FSLQ) as well as the sliding mode controller for the tracking of the human replicated lane-changing trajectory \cite{chee1994lane}. Wang et al. determine optimal lane change times and accelerations by minimizing an objective function, which considers driving safety, efficiency and comfort criteria for connected and automated vehicles. They included driver comfort  by penalizing large accelerations or decelerations for strategic overtaking and cooperative merging scenarios \cite{wang2015human}. %??use one sentence to define these new terms.??
\EditFL{ Hatipolglu et al. presented a nonlinear controller to track the desired yaw angle with consideration of the vehicle dynamics and the actuator dynamics \cite{hatipolglu1997steering}.} In addition, both radar and camera are used to detect the road curvature. Later, Taylor et al. used a vision based lateral control system to investigate system parameters, such as vehicle velocity, how far a vision sensor can see, and computational time related to the control systems of a vehicle \cite{taylor1999comparative}. They tested three feedback control strategies on the lateral control task with an experimental vehicle. These strategies show acceptable performance in terms of replicating lane-changing comfort on the straight and curved roadway sections. Keviczky et al. used model predictive control (MPC) to minimize the tracking errors as well as the control input for the lane-changing behavior \cite{keviczky2006predictive}. The experimental simulation result shows that the MPC controller can have a good stability performance under the high velocity. {\major The `Minimizing Overall Braking Induced by Lane Changes' or MOBIL lane-changing model~\cite{kesting2007general,treiber2016mobil} %, is based on strategic features of classical game theory. The strategic features are described using a politeness factor, which can vary different human driving behaviors. This model consists of two criteria: incentive and safety. Based on this model, the incentive criterion will be used for determining the attractiveness of a target lane for a subject vehicle. If the incentive criterion is adequate, the safety criterion needs to check to make a lane change. The incentive criterion is measured by determining utility (weighing the advantage to change a lane against disadvantage, which is imposed by other vehicles) of a target lane. According to safety criterion, safety for changing a lane is measured by the risk associated to change the lane acceleration.
applied strategic features of the classical game theory. The strategic features are described using a politeness factor, which can vary depending on different human driving behaviors, such as altruistic, realistic, selfish and malicious driving behaviors. This model consists of two criteria: incentive and safety. The incentive criterion determines the attractiveness of a target lane for a subject vehicle. The incentive criterion is measured by determining utility (weighing the advantage to change a lane against disadvantage, which is imposed by other vehicles) of a target lane. If the incentive criterion is satisfied, the safety criterion needs to be checked before performing a lane change. According to safety criterion, safety during lane changing is measured by the risk associated to change a lane (i.e., acceleration).} Naranjo el. al. adopted Fuzzy logic based controller to imitate the human decision about when to manipulate the lane changing behavior based on the headway and velocity difference between {\acpt a} subject vehicle, and preceding and following vehicles in a target lane \cite{naranjo2008lane}. The fuzzy lateral movement controller captures human driving behavior using the experts' procedural knowledge related to the lane-changing behavior.

%\subsubsubsection{Trust}
%\vspace{0.01in}
\noindent \textbf{Trust.} To realize full benefit of the CAVs, automated system must earn human trust so that users can rely on the system. Any automated system needs a high level of trust to mass adoption of the technology \cite{weinstock2012effect}. %Also, if users do not have enough trust, it is not possible to take full advantage of the {\acpt automated} system.
The trust of an automated system can be measured with the system accuracy \cite{lees2007influence} and reliability \cite{blanco2015human}. Lee and Seppelt found that high false alarm rate (i.e., accuracy) decreases the system reliability and compliance of an  automated system \cite{lee2009human}. Moreover, Seppelt and Lee found that it is more effective to provide continuous information to the users regarding the state of an automated vehicle instead of providing immediate warning because of system failures \cite{seppelt2009supporting}. {\blue One needs to understand trust factors for providing guidance to the CAV system developers}. Carlson et al. identified twenty-nine factors that can compromise trust of CAV user, and performed statistical analyses for automated vehicle related factors and trust factors related to the safety features of a vehicle \cite{carlson2014identifying}. The critical factors identified for user's desirability and reliability of the automated cars are: i) level of accuracy of the vehicle's routes; ii) availability of current roadway information (e.g., weather, traffic congestion, and construction) to a vehicle; iii) level of training and prior learning of a vehicle; iv) system failure detection (e.g., making a wrong turn, running a stop light); v) accuracy of the route selection; vi) user's familiarity with the vehicle features; vii) agreement of routes between vehicle and user's knowledge; viii) the vehicle's methods of information collection \cite{carlson2014identifying}. %V2V and V2I communication can provide more information regarding surrounding environment to a CAV to provide desirable increase the trust of the CAV system.

\vspace{0.05in}\subsubsection{Adaption to the Designed CAV System by User}
Behavioral adaptation and situational awareness are %the key factors, and
summarized in the following sub-sections.

\vspace{0.02in}
\noindent \textbf{Behavioral adaptation.} In CAV system, Human-Machine Interface (HMI) plays a critical role as the HMI assists user to change user's role from an actuator to a supervisor or vice-versa \cite{cunningham2015autonomous, baquero2009human, martens2008human, toffetti2009citymobil}. A user needs to adapt to the HMI interface of a CAV to execute appropriate decisions through voice command, touch or any other haptic (i.e., gesture) command. It is warranted to increase CAV user education about the system functionality of an automated vehicle \cite{larsson2010issues}. If users are more knowledgeable about the system and their limitations, users will be more aware of such system and they will adapt to the system \cite{larsson2012driver}. In addition, a user could act as a sensor in a CAV system and could provide input to the system controller depending on the different driving scenarios (e.g., congested/uncongested roadway traffic condition, merging or diverging traffic scenario). Recently, gesture-based automated interface has been explored using different sensor technologies for vehicle control, primarily for automated vehicles, along with voice and touch interface \cite{fong2001advanced}. When an occupant or a user is unable to interact with the CAV system through voice or touch interface, a gesture-controlled system could be very effective \cite{cunningham2015autonomous}. A CAV vehicle user performs a gesture (e.g., the motion of hand), and the CAV can interpret and react in a manner that is commensurate with the users' intentions.

%\vspace{0.1in}
\noindent \textbf{Situational awareness.} Situational Awareness can be defined as the awareness of a user regarding the surrounding environments in a CAV system. It has been suggested that automation may lead to users not informed of the surrounding situations and hence loss of situational awareness
\cite{parasuraman1993performance}. Endsley defines situational awareness as user's constant attention, on events that are going on around, in a dynamic human decision-making environment, and based on the current information one also needs to forecast near future events \cite{endsley1995toward}. If we consider a CAV system, a user needs to be aware of the surrounding dynamic environment and needs to perform an action based on an extreme emergency (e.g., system failure of a CAV). According to Endsley, situational awareness of a dynamic human decision-making environment can be divided into three levels: (1) perception of a situation, (2) comprehension of a situation, and (3) prediction of the future states based on the comprehension of a current situation \cite{endsley1995toward}. A well designed CAV system must be able to provide information to the user in a regular interval to take any action in a timely manner on a critical situation (e.g., system failures). Two categories of a situational awareness system for a CAV are : i) engagement and disengagement; and ii) mode confusion.
As a user of a CAV system, a passive fatigue (e.g., decreased driving task engagement) may occur due to engaging and disengaging with this system. Such sudden shifts in vehicle operation can require long reaction time during safety-critical driving events, such as roadway incidents. The effect of automation on driving behavior as it relates to how a user in a CAV environment will react during engaging and disengaging, and procedures to facilitate reliable and safe transition are required to design a CAV with high reliability. Because of automation (workload reduction), CAV users will engage in non-driving related tasks that can distract a user from the supervising role, which will lead to risky situations in case of system failures and emergencies. On the other hand, mode confusion is a phenomenon that can be defined as a discrepancy between the driver expectation from a designed CAV to operate the system and the actual operation procedure of a CAV \cite{cummings2014shared} \cite{martens2013road}. If a user of a CAV is not aware of the state of the vehicle, a user could make decisions based on the certain belief, which may not be correct \cite{bredereke2002rigorous}. In next sections, we discuss the possible current challenges and future research directions on three key factors of CAV systems.

%\vspace{0.1in}
\subsection{Challenges}
%\vspace{0.1in}
%\noindent \textbf{:}
{\major In this section, we discuss the challenges related to human factors of CAV systems as follows.

\vspace{0.05in}
\noindent \subsubsection{User Preferences Modelling}

The biggest challenges for modelling the longitudinal motion (i.e., car-following behavior) and lateral motion control (i.e., lane changing and overtaking) considering human preferences are to detect the environment and predict the intentions of the neighbor vehicles. The sensing capability of a sensor must be very accurate if the distance between two vehicles is very short \cite{levinson2010robust}. Moreover, since the merging action is a complex scenario because of the interaction between the subject vehicle and the nearby  vehicles in the target lane, it is crucial to know the intention of the neighbor vehicles. Any failures in these two aspects will lead to a catastrophic result, such as the instability of the vehicle dynamics \cite{gipps1986model}. Although, in dealing with high crash risk, a conservative algorithm should be developed to deal with the uncertainties, which will heavily deteriorate the efficiency of the lateral motion generation, and user comfort and acceptance \cite{best2017autonovi}. The emergence of V2V and V2I communication technologies (discussed in Section \ref{sec:sensingComminication}) can help to solve these uncertainties as well as the inefficiency \cite{yang2004vehicle}.

%It is already shown using the simulation that the longitudinal motion of each vehicle in a CACC platoon, which uses communication among the vehicles, will increase the performance of the controllers in terms of user safety, user comfort, and traffic efficiency aspects (i.e., the capacity of a roadway) \cite{wang2015human}. However, in more complex scenarios, such as the intersection lane-changing behavior, the lateral controller with V2V and V2I functions obviously outperform the conventional prediction based controller.

\vspace{0.05in}
\noindent \subsubsection {CAV's System Failure}
To study the effects of stimulus-independent thought, which could occur before, during and/or after a transition, it is necessary to develop a suitable method to establish what mechanisms contribute to the situational awareness to a CAV user in case of system failures. Surrounding Information of a CAV through V2V and V2I communication needs to use to provide early warning to the user for an unsafe situation.

%\FL{??last sentence is not clear enough}

\vspace{0.05in}
\noindent \subsubsection{HMI}
A CAV system must have an HMI, which should replicate and produce repeatable results. With advance in vehicle technology, there is a need for in-depth study how human will interact with automation features, such as how to minimize user-introduced errors, consequences of over or under relying on the system, and effectiveness of different user feedback system interfaces and design of an accurate human-centered controller by obtaining feedback through HMI.
%{\major ??only gesture controller is very narrowed field of HMI. There are a lot of other factors (views, option menus, and so on) to be considered for well-designed human machine interactions. Maybe, no need to start with Gestures. Does HMI have some associated with CAV controller? }

\vspace{0.02 in}
\subsection{Future Research Directions}
{\major A primary goal of the most automation is to achieve a high reliability. For the CAV system to be acceptable, user preferences must be met. In previous discussion of current state-of-the-art understanding of human factor issues, research trends and challenges are discussed human factors related to a CAV. In this sub-section, we identify future research directions on major human factor issues, which include: i) CAV user preferences modeling using artificial intelligence; ii) Engaging in the case of CAV's system failure; and iii) Integration of multiple assistance systems in the HMI.

\vspace{0.02in}
\noindent \subsubsection{CAV User Preferences Modelling using Artificial Intelligence} User in each vehicle of a CAV will have different preferences, such as preferred speed, gap between vehicles, in different driving condition \cite{kuderer2015learning}. Thus, a CAV must allow different driving preferences (e.g., speed, acceleration) based on the user characteristics such as age (young, older) and gender (male or female). One of the major challenges of the CAV system is how to incorporate these user preferences in the real-time CAV operation. A machine learning-based model can be developed for human-centered CAV speed recommender system depending on the human behaviors. It is possible to train different user behaviors with collected vehicle trajectory data for a comfortable speed, acceleration, and the gap from the immediate front vehicles. In addition, the modeling of lane changing (merging and diverging) behaviors depends on the modeling accuracy of user preferences, such as user gap acceptance behavior, and available gap for changing lane in the target lane. Thus, it is required to develop a user-oriented merging and diverging models by incorporating human preferences for the CAV system.
\vspace{0.02in}
\noindent \subsubsection {Engaging in a Case of CAV's System Failure} CAV users will only require engaging in a supervisory role if there is a safety-critical situation in the case of CAV system failures. However, users' ability to do so is limited by humans' capacity for staying alerted when disengaged from the driving task. Manufacturers and other entities need to incorporate {\blue users} engagement monitoring system for a CAV system. Therefore, a suitable method, such as integration of different warning system (e.g., visual, audible and vibration warnings), needs to develop based on the information from the controller to increase awareness of a critical situation so that CAV users can be engaged in the system to control the situation.
%{\major ??connection with controller is not so obvious}

\vspace{0.02in}
\noindent \subsubsection{Integration of Multiple Assistance Systems in the HMI} HMI plays a critical role to inform and take an appropriate decision through touch and voice command or any other gesture command. Thus, it is important to study in-depth how human will interact with automation features. At a minimum, HMI interfaces of a CAV system should be capable of functioning reliably and providing accurate information (e.g., a malfunction of the CAV system) to the user \cite{FederalVehiclePolicy}. We also need to investigate how we can integrate multiple assistance systems in the HMI so that we can ensure the reliability of the system.}

%\vspace{0.1in}
\section{Information-aware Controller Design}
%\vspace{0.1in}
\label{sec:controls}

The AV controller is expected to generate the appropriate control command at anytime such that automated driving tasks can be fulfilled. Due to the environmental uncertainties caused by the  limitations of sensing devices, the overall efficiency of the sensor-based AV  might be  compromised to guarantee the driving safety. However, with the help of extra information from the V2X communication, the corresponding information-aware controller should achieve better driving performance than the conventional AV controller. In this section,  we focus on how the controller utilizes the information through the wireless communication.

In Section~\ref{subsec:c1}, the current status of the V2X information utilization is summarized. The typical techniques to realize the algorithms are also mentioned. The current challenges of the information-aware controller are discussed in Section~\ref{subsec:c2}. Finally, the future research directions are presented in Section~\ref{subsec:c3}.}

%\vspace{0.1in}
\subsection{Current Status}
%\vspace{0.1in}
\label{subsec:c1}

\begin{figure} [H]
\vspace{-0.1in}
\includegraphics[scale=0.3]{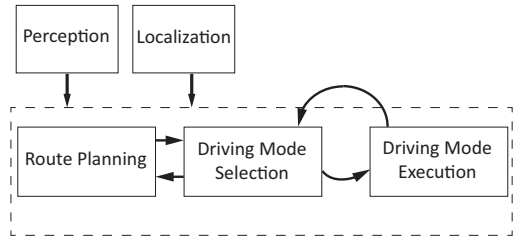}
\centering
\vspace{-0.05in}
\caption{Controller block diagram.}
\label{fig:block_diragm}
\vspace{-0.1in}
\end{figure}

The functional structure for an AV controller with Level 5 automation, according to the SAE classification~\cite{sae2016taxonomy}, can be composed of five layers: perception layer, localization layer, route planning layer, driving mode selection layer, and driving mode execution layer~\cite{milanes2010clavileno}~\cite{okuda2014survey}. These five layers are shown in Fig. \ref{fig:block_diragm}. The main task of the perception layer is to perceive the environment based on the information collected by different \FL{sensing and communication technologies, which are summarized in \plag{Section}~\ref{sec:sensingComminication}}. \EditFM{Localization layer is required to locate the position of a subject AV on a given map.} The route planning layer \EditFM{is responsible for an} optimal route from an original to a destination. Moreover, the driving mode selection layer \EditFM{is developed to} determine  which driving mode should be chosen under current driving situation. \FL{For example, if the preceding vehicle is too slow, the current vehicle-following mode can be switched to overtaking mode (lane-changing mode) when all the safety-relevant criteria are satisfied}. \EditFM{Finally,} the chosen driving mode is executed under the cooperation among sensing systems, control algorithms, and actuators. \EditFL{It has been demonstrated that the information from on-board sensors and GPS  is enough \EditFM{for the realization} of an automated driving task~\cite{levinson2011towards}.} However, the emergence of V2X communication can provide the vehicle with additional information beyond what the on-board sensors can offer, which can further improve the CAV driving efficiency and safety~\cite{nagel2007intelligent}.
{\blue Among the five layers of the controller, perception and localization layers are more relevant to the  information collection and processing tasks, discussion of which are beyond the scope of this \EditFL{review paper}. In the following paragraphs, we focus on the information-aware controller design \EditFM{in terms of} the other \EditFM{three} different layers and discuss how V2X communication can enhance the controller design.
%\begin{figure}[h]
 %%%   \hfill
  %\subfloat[With communication (centralized)]{\includegraphics[width=0.24\textwidth]{plan_3.eps}\label{fig:a3}}
%  \vspace{-0.05in}
  %\caption{Route planning comparison.}
 % \label{fig:a}
%  \vspace{-0.1in}
%\end{figure}
%{\acpt [??????   THIS FIGURE DOES NOT DISTINGUISH DISTRIBUTED AND CENTRALIZED SYSTEM CLEARLY.......THIS FIGURE SHOULD BE SELF EXPLANATORY AND TEXTS RELATED TO THIS FIGURE SHOULD EXPLAIN THE WHOLE PROCESS AND EXPLAIN EACH TERM MENTIONED IN THIS FIGURE ....... I WILL CALL AND EXPLAIN YOU WHAT ARE THE PROBLEMS...... ?????] }
%\vspace{0.1in}
\subsubsection{Route Planning Layer }
%\FL{ Figure 4 should be discussed clearly, maybe one more line at the beginning or at the end**solved**}

The route planning layer develops an optimal route from an origin to a trip destination based on various criteria, such as shortest distance, \EditFM{shortest traveling time}, and avoidance of tollways. \FL{The corresponding path searching/planning algorithms have been well-developed in recent few decades~\cite{pugliese2013survey}}. For an AV that fully relies on the on-board sensors, the optimization of the route merely depends  on the information in the given map, such as route distance and tollways information~\cite{leshed2008car}. In this case, \EditFL{the shortest traveling time} is hard to achieve since there is no information about the \EditFM{real-time} traffic conditions, such as \EditFL{roadway traffic congestion condition and incidents}. {\blue When the sensor-based AVs encounter traffic congestion or incidents, alternative routes planning will be adapted, which may not be efficient in terms of energy consumption and travel time}~\cite{kammel2008team,bacha2008odin}. \FL{The details about how V2X communication can enhance the route planning layer are discussed in the following paragraph.}

To solve the inefficiency caused by limited information, V2X communication technology can be adopted to provide real-time traffic conditions to the CAV controllers~\cite{fontanelli2010dynamic}, where distributed control scheme determines the optimal route individually for each CAV. The chosen route can be optimized initially with respect to the travel time. In addition, an optimal route planning from a centralized controller can also be realized via the adoption of V2X communication, where the optimal routes of different CAVs are calculated by considering their impacts on the overall traffic condition~\cite{pan2012proactive,wang2015real,de2016real}.

~\EditFM{\FL{In the work}~\cite{fontanelli2010dynamic}, motion states, i.e., velocity and acceleration, of different vehicles are gathered in the cloud via V2I communication. Then, the traffic simulation software AIMSUN~\cite{choffnes2005integrated} is utilized to predict the future traveling time for each roadway segment. Finally, the fastest route is derived by minimizing the overall travel time via the modified Dijkstra's algorithm~\cite{skiena1990dijkstra} where the cost of each segment is the dynamically predicted travel time.} However, this distributed optimization for each individual CAV might switch the road congestion from one spot to another. \EditFL{To deal with this issue, \FL{Pan et al. presented} the Entropy Balanced \textit{k} Shortest Paths strategy~\cite{pan2012proactive}, where the route selections of all CAVs on the road are managed by a centralized controller via V2I communication.} On the top of the Random \textit{k} Shortest Paths (RkSP) method~\cite{jimenez1999computing}, the optimal routes are assigned upon the driving task urgency of CAVs. The vehicles with higher urgency are assigned to the optimal routes first. Using the BSMs from each vehicle via V2I communication, the roadway bottleneck effects can be predicted based on the calculation of  traffic density}~\cite{backfrieder2017increased}. In combination with the A* searching algorithm, which is an classical method to find the general route with the lowest cost, e.g., shortest travel distance~\cite{stentz1994optimal}, it is shown that the presented route planning algorithm can efficiently avoid the roadway bottleneck. In the work~\cite{wang2015real}, besides the route length, the real-time traffic congestion information is included in building the cost function for each route segment. As a result, the route is optimized based on both travel time and travel distance aspects. In addition, to realize an optimal route selection, the dynamics of the roadway intersections are incorporated to predict the traffic dynamics under different route selections. \FL{The minimization of the corresponding cost function is solved by Lyapunov optimization process~\cite{zheng2014distributed}}. By considering the \EditFL{limited V2I communication coverage issue of RSU}, \FL{Souza et al. developed} a novel route re-planning algorithm~\cite{de2016real}. As the coverage of the RSU is considered, the centralized controller for route planning only assigns the optimal routes to the CAVs within the RSU covered area without changing the ending route point on the edge. For the uncovered area, the CAV follows the original plan initiated from \FL{the boundary of V2I covered area}.
%\FL{Similarly, RkSP method is used to find the optimal path where Boltzmann probability distribution method is adopted to avoid the {\acpt congestion spot switch problem [WHAT IS "CONGESTION SPOT SWITCH PROBLEM" ????...EXPLAIN IT IN PLAIN ENGLISH SO THAT EVERY READER CAN UNDERSTAND EASILY ]}~\cite{gardiner2009stochastic}.}

\vspace{0.05in}
\subsubsection{Driving Mode Selection Layer}
Driving mode selection layer determines which driving mode should be chosen based on the knowledge of the {\blue roadway traffic conditions}, such as the position and velocity information of neighbor vehicles. There are numerous driving modes dealing with different situations, such as vehicle-following mode, lane-changing mode(e.g., merging or, leaving from a lane), avoiding obstacle mode, and parking mode. The conventional AV with the on-board sensors can still realize the driving mode selection task successfully~\cite{urmson2008autonomous}.  However, it cannot identify the intention or future motion  states of its neighbor vehicles directly which is essential for the selection of the driving mode. Although \EditFM{learning-based} algorithms can be used to predict such information  (e.g., acceleration levels in the next few time steps)~\cite{ferguson2008motion}, the driving efficiency might be sacrificed when dealing with uncertainty of the neighbor vehicles' driving trajectories. %Generally, additional information via communication can contribute to a reasonable driving mode selection, which will be explained in the next paragraph. However, some of the driving modes are determined by the chosen route, such as the intersection and merging/diverging sections.
In the following paragraph, we summarize the scenarios where the AV has the freedom to choose the driving mode, such as obstacle avoidance mode.

For the AV without wireless communication capability, the distance and speed information of itself and its neighbor vehicles within the detection area of the sensors is mainly used to design the control algorithm. In~\cite{kim2009design}, the inter-vehicle states are classified into three regions (i.e., region I,  region II, and region III) based on the relative velocity and headway between the vehicle and its preceding vehicle/obstacle. If the rear-end collision can be avoided by a mild deceleration (region I), \EditFM{the vehicle will continue following the vehicle in front of it by applying the brake.} If the desired deceleration is too large, the lane-changing (region II) or both deceleration and lane-changing (region III) should be applied. \FL{Kala et al.} presented a priority-based algorithm in~\cite{kala2013motion}, where multiple modes can be activated simultaneously by the low-level logic, e.g., lane-changing mode and obstacle avoidance mode are both triggered. To deal with this issue, the priorities have been assigned to different driving modes to prevent the potential mode conflicts. In the above case, for example,  the safety-relevant obstacle avoidance has a higher priority. %In addition, only when the \EditFM{corresponding safety-relevant constraints} are satisfied can the driving mode be executed.
%For the lane-changing mode, prevention of collision should be the constraint.
\FL{The details about how wireless communication can improve the design of the driving mode selection layer are provided in the following  paragraphs.}

\vspace{0.01in}
%\noindent\textbf{Information-aware driving mode selection.}~

\EditFL{The communication technologies can provide accurate and \EditFM{prompt} vehicle trajectory information without the constraints brought by the on-board sensors of a CAV.
The \EditFM{algorithm} presented in~\cite{andrews2012vehicle} utilizes the V2V communication \EditFM{technology} to derive relative speed and distance with respect to the preceding vehicle. \EditAA{As can be expected, this \EditFM{algorithm} has a fast and accurate decision in whether to adjust the speed or conduct a lane-changing maneuver to deal with the slow front car/obstacle.}} \EditFL{Besides the advantage of the higher quality and extra information, V2V \EditFM{communication} enables the \EditFM{negotiation} between CAVs to improve the decision-making process.} In~\cite{caveney2012cooperative}, the pre-calculated trajectory of the CAV is shared with its neighbor vehicles. \FL{The pre-defined trajectory is then used as a reference to generate collision-free trajectories for other vehicles, which can be solved by the standard terminal-constraint optimization method~\cite{geroliminis2013optimal}.} This pre-defined control command sharing will surely enhance the effectiveness of the CAV's decision-making. For example, when a slower vehicle is detected ahead of an AV on the left lane (considering {\acpt two lanes} in each direction), the AV on the left lane only with on-board sensors cannot determine whether the slow-moving vehicle in front of it will change to the right lane within a next few seconds or not. However, with the negotiation and information sharing between the subject CAV and the CAV in front of it, this dilemma can be eliminated.

In addition, V2I communication enables {\blue a} centralized controller to collect motion states information of all {\blue CAVs}. Hence, the centralized traffic management controller can realize a global optimization. Under this scenario, the driving mode selection controller of CAV is assumed to follow the instructions from a traffic management center. Based on the assumption that traffic information is known via V2I communication, \FL{Roncoli et al.} developed a centralized traffic flow controller to optimize the traffic flow rate~\cite{roncoli2015traffic}. In this centralized controller, a macroscopic traffic flow model is adopted where traffic segment speed, lateral (lane-changing) flow, and the freeway ramp entering rate are the control inputs. \FL{All these three control inputs are calculated optimally through minimizing the quadratic traffic congestion cost function which can be efficiently solved by the optimization solver~\cite{gurobi2015gurobi}.} Moreover, the  flow rate control signal is then desegregated into the individual vehicle controller command to the CAVs. \FL{In the work}~\cite{roncoli2015model}, an MPC controller is developed to deal with a similar traffic flow optimization task, but considering the existence of {\blue manually driving vehicles,} where the manual driving flow model is calibrated by the traffic pattern observation. In addition, the control input of the traffic dynamics becomes the manipulation of CAVs instead of all the vehicles in the mixed traffic flow.

\vspace{0.1in}
\subsubsection{Driving Mode Execution Layer}
\vspace{0.01in}
In the following section, we will introduce how wireless communication can help to execute appropriate driving modes.

\vspace{0.01in}
\noindent\textbf{Vehicle-following mode.}~For the AV without communication capability, the automated vehicle-following mode is called an adaptive cruise control (ACC) system, which is used to maintain a reasonable headway between the front vehicle and the following vehicle~\cite{winner1996adaptive}.There are numerous criteria to define the desired headway, such as safety and human factors. Safety criterion guarantees that no rear-end collision occurs during the vehicle-following scenario. Human factors of CAV are covered in Section \ref{sec:human}. ACC system mainly collects the headway and velocity information as the input of the controller to determine the desired acceleration. To regulate the tracking performance, there are two stability requirements~\cite{liang2000string}: individual stability and string stability. For the individual stability, it requires that AV should track the desired headway successfully. For the string stability, it requires that the fluctuation of the motion state should not propagate upstream along the platoon. To achieve the string stability, the definition of the desired headway is critical. Moreover, the constant time headway is a widely adopted spacing policy of a platoon of AVs to define the desired space headway as in~\cite{rajamani2002semi}. The desired space headway, $D$, between the subject and the preceding vehicles is as follows~\cite{rajamani2002semi}:
\begin{gather}
\label{eq:traffic_capacity}
D = h\cdot v+d_{min}
\nonumber
\end{gather}
where, $h$ is the constant time headway, $v$ is the subject vehicle velocity, and $d_{min}$ is the minimum inter-distance between the subject and the immediate preceding vehicle.

In general, a typical proportional-integral-derivative (PID) controller~\cite{martinez2007safe,ioannou1994time} or a sliding mode controller~\cite{xiao2011practical} can minimize the headway error and  guarantee a string stable ACC design even under the effect of the powertrain dynamic delay.

When the information from V2X communication is available in the vehicle-following function, the controller is enhanced as the CACC system.  Since the V2V and V2I communication technologies are used, there is freedom for the controller to utilize various information from the platoon with less limitation on the information type and relative position. Currently, several communication typologies have been investigated including the predecessor-following (PF) topology, the bidirectional (BD) topology, the predecessor-following leader (PFL) topology, and the two predecessor-following (TPF) topology~\cite{li2015overview}. \FL{Naus et al. adopted} feedforward controller with communication information in combination with the conventional ACC feedback controller~\cite{naus2010string}.  It is found that CACC system can realize a shorter time headway than the ACC system to guarantee the string stability. In addition, this CACC system controller structure has a better performance in minimizing the velocity fluctuation for upstream traffic flow \EditFM{with the help of feedforward controller enabled by V2V communication}~\cite{ploeg2014lp}. \EditAA{Moreover, V2X communication can enable the controller to deal with new challenges via advanced control algorithms.} The reinforcement learning approach is adopted in~\cite{desjardins2011cooperative} to build the controller. \FL{Control inputs are learned from the CACC simulator with high-fidelity nonlinear vehicle dynamics where more rewards will be received if the vehicle inter-distance is close to the desired one.} The simulation shows that this novel CACC controller can guarantee an efficient tracking of the desired space headway. In~\cite{kamal2014smart}, an MPC controller is developed that includes a deviation between a calculated headway and a desired headway of a following vehicle, and  the speed fluctuations of the following vehicle caused by the deviation in  the cost function of the MPC controller. When an MPC controller is adopted, different objectives can be realized by customizing the computationally feasible cost function. \EditFL{For example, jerk minimization can be considered in the cost function with an objective to enhance a CACC occupant's} comfort~\cite{wang2015human}.

\vspace{0.01in}
\noindent\textbf{Lane-changing mode.}~ Typically, the lateral controller (also known as a lane-changing controller) of an AV is composed of trajectory generation task and trajectory tracking task to change lane from the current lane to the target lane. It is critical for the lane-changing controller to make sure that no collision will occur when an AV changes lane from one lane to another lane. At a high level, an AV lateral controller generates position and velocity trajectory to perform the lane-changing task with consideration of different factors, such as safety, comfort, and traffic flow efficiency~\cite{hatipoglu2003automated}. At a low level, a trajectory tracking controller of an AV tracks the position and velocity trajectory smoothly and accurately~\cite{hedrick1994control}.  In this section, the lateral trajectory generation task with V2V communication is discussed where communication can increase the safety, comfort, and traffic flow efficiency.

To guarantee the safety, Davis et al. concluded that the headway between the subject vehicle,  which is intended to change from the current lane to the target lane, and the vehicle on the target lane, should be a function of the subject vehicle velocity and the velocity difference between the subject vehicle and the vehicle in front of it~\cite{davis2007effect}.  In~\cite{lu2004automated}, the velocity trajectory generation algorithm  is developed for an AV, which merges into the automated vehicle platoon. At the end of the merging process, the algorithm ensures that the velocity of the subject vehicle is equal to the platoon velocity and the space headway is matched with the desired space headway of the platoon. Moreover, it ensures to follow the minimum space headway during the merging process to avoid collision with the preceding vehicle and to guarantee user comfort. However, maximum acceleration and deceleration of the neighbor vehicles are usually considered in the existing lane-changing algorithm to calculate the collision-free trajectories that compromise driving efficiency~\cite{jula2000collision}. In addition, the velocity fluctuation is almost inevitably introduced that decreases the riding comfort if a vehicle suddenly merges into the target lane without any trajectory negotiation with vehicles in the current and target lane~\cite{davis2007effect}.

In the research~\cite{marinescu2012ramp}, a slot-based merging algorithm is presented with the utilization of V2V communication. When a vehicle intends to merge onto the target lane from the current lane, the slot-based traffic management system (TMS) of the subject vehicle will find an empty slot for the vehicle after negotiation with the neighbor vehicles or the subject vehicle will receive a rejection of the request in case of slot unavailability. According to the slot-based method, all the vehicles on a roadway move based on the availability of the empty slots. The TMS can make a slot available for a subject vehicle by negotiating the movement with the surrounding vehicles. By adopting V2V communication, safety can be guaranteed without using the conservative mathematical constraints, e.g., assuming the maximum acceleration value will be adopted by neighbor vehicles. Moreover, the vehicles in the target lane have more time to prepare for lane changing, which can minimize the velocity fluctuation.

Wang et al. developed a proactive merging strategy of a vehicle to improve merging efficiency ~\cite{wang2007proactive}. The priorities of the vehicles on the target lane are determined by the distance between the current location and the merging point. The vehicles with lower priorities should adjust their velocities accordingly to leave enough gap for the merging vehicle based on the instructions from RSUs. Hence, the subject vehicle can directly merge into a target lane on the freeway/highway from the ramp  smoothly. As the velocities and inter-distances of the involved vehicles are  harmonized beforehand, this pre-computed trajectory negotiation mechanism can eventually be used to realize a fluctuation-free lane changing action.

%\FL{what is reference vehicle?? what is host vehicle in above paragraph??**solved**}

%\vspace{0.01in}
\noindent \textbf{Intersection mode.}~For an AV without communication capability, it will follow the {\blue roadway} traffic control rules, such as the stop sign and traffic signals when driving under the intersection mode~\cite{franke1998autonomous}. However, the V2X communication technology can offer a more efficient and safer roadway intersection mode execution.

\EditFL{Firstly, V2I communication  transmits the traffic phasing and timing (SPaT) information to CAVs. Hence, the CAVs can utilize the SPaT information to {\blue operate} the controller for intersection mode. \FL{Malakorn et al. calculated} the constant  vehicle acceleration value and duration to avoid the red traffic {\blue light indication}~\cite{malakorn2010assessment}. It was shown that this calculation can significantly improve the overall fuel economy and traffic flow throughput. Moreover, a roadway corridor with multiple traffic signals are considered in~\cite{asadi2011predictive} to calculate the desired velocity range.} When a CAV passes the first traffic signal within a specific velocity range, the CAV controller will continue checking the feasibility of passing the next traffic signal before it turning into red light. The final velocity range is determined \newc{when the CAV will inevitably stop at the next intersection} i.e., \EditFM{there is no feasible velocity range to pass all the traffic signals}.

\EditFL{Secondly, the V2V communication is more helpful at an intersection without signals, where the efficiency is heavily deteriorated due to the rigid traffic stop sign law.} In~\cite{milanes2010controller}, the V2V communication is used to broadcast the position and velocity information to  other vehicles within the coverage area of the unsignalized intersection. To travel through the unsignalized intersection, the AV should yield to the vehicles with \FL{higher priorities} based on the traffic rule.  A fuzzy logic controller is adopted to control the level of the throttle and brake based on the space headway and inter-velocity with respect to the higher-priority vehicles. \EditFL{Moreover, an intersection control agent (ICA) is designed~\cite{lee2012development} as a centralized controller to manage the vehicles at the  intersections via V2I communication.} The trajectory of each vehicle is managed to make sure that there is no trajectory overlaps to prevent the collision. \FL{The velocity trajectories of the vehicles are solved by nonlinear constrained programming (NCP)~\cite{wright1999numerical} with the consideration of minimum time headway constraints and maximum level of acceleration constraint.} If the trajectory overlap is inevitable, the vehicles with lower priorities in the queue will be stopped. In addition, the corresponding stop-recovery algorithm is designed to guide the stopped vehicle in the intersection scenario.

\EditFL{From the above review, it is clear that V2X communication can enhance the performance of CAV controller since more information can be used to improve the controller efficiency and the performance in dealing with the uncertainty {\blue of the roadway traffic condition}. In addition, V2I communication can further contribute to \newc{developing} a centralized controller, where global optimization can enhance the overall traffic condition.}

%\vspace{0.02in}
\subsection{Challenges}
%\vspace{0.1in}
\label{subsec:c2}
    \FL{The information-aware controller for CAVs has drawn a fair amount of attention. In addition, several applications have been well developed, such as the CACC, routing planning and intersection mode controller. However, the followings are the challenges that need to be addressed to develop information-aware controller considering communication and human factors issues.}

%\vspace{0.1in}

%\vspace{0.1in}
\subsubsection{Centralized versus Distributed Controller}
\vspace{0.01in}

%\noindent \textbf{Centralized versus distributed controller.}~
\FL{It has been demonstrated that the centralized controller can maximize the overall benefit in terms of traffic flow efficiency  by assigning control commands to multiple vehicles} in a coordinate fashion ~\cite{de2016real}. \EditAA{However, the control authority issue of each individual vehicle is still significant with {\blue respect} to the liability issue~\cite{duffy2013sit}. In addition, high communication bandwidth and computation capabilities are required in the realization of a fully centralized controller of CAVs,  which is still difficult to achieve in the near future. Meanwhile}, the distributed computation device in the individual CAV might be incapable of processing all the information from the communication path in real-time. As a result, how to make a trade-off between a centralized control structure and distributed control scheme should be addressed.

%\vspace{0.1in}

%\vspace{0.1in}
\subsubsection{Communication Imperfection}
\vspace{0.01in}
%\noindent \textbf{Communication imperfection.}~

\FL{ Both the centralized control and distributed control of the AV can have plenty of information to  execute the task, such as vehicle motion stabilization and control command optimization via V2X communication. However, most of the researchers assumed perfect communication scenario without any information loss ~\cite{wang2015human}, which is unrealistic and hence the results are not reliable. Recently, the effects of communication imperfection on the longitudinal speed control have been studied ~\cite{naus2010string,xu2002effects}. The results suggest that communication delay, shadow fading, and interference effects can heavily deteriorate the performance of the vehicle following mode execution in terms of safety, traffic efficiency, and driving comfort. However, the communication imperfections on other layers (i.e., route planning and driving mode selection) has not been well studied in the existing literature.}

%\vspace{0.1in}
\vspace{0.1in}
\subsubsection{Mixed Traffic Scenario}
\vspace{0.01in}

%\noindent \textbf{Mixed traffic scenario.}~

\FL{The existence of the vehicles without any communication capability in the real traffic scenarios will challenge the efficiency of CAV controllers. Without V2I communication capability, the vehicles cannot broadcast the motion states information to the centralized controller. Moreover, the control commands of the manual driving vehicles are not available for sharing.} Hence, the calculated trajectory from the centralized controller will be deteriorated. The V2V-based distributed cooperative controller might also need to be degraded to the sensor-based controller, which could compromise roadway efficiency. As CAVs will penetrate the market gradually~\cite{specks2005car}, the coexistence of the CAVs and non-connected vehicles should be expected and considered in the designs of intelligent controllers.

%\vspace{0.01in}
\subsection{Future Research Directions}
%\vspace{0.05in}
\label{subsec:c3}
\FL{Although the current automated car controllers work well to fulfill various control objectives, there is still much room left for further research. In the following paragraph, major controller challenges in the information-aware controller design are presented to show the potential directions in future work.}

%\vspace{0.05in}

%\vspace{0.1in}
\subsubsection{Hybrid Structure with Centralized and Distributed Control Scheme}
\vspace{0.01in}

%\noindent \textbf{Hybrid structure with centralized and distributed control scheme.}~

It can be expected that the V2I-based centralized controller, which is located at  a traffic management center, will play an important role in the future ITS. The CAVs can also have an on-board unit with high computation capability. Hence, it is very important to incorporate the centralized controller and distributed controller via V2X communication. There are several papers taking advantages of the V2I communication on top of the typical V2V-based controller designs in terms of the higher computation capability~\cite{bento2012intelligent} and richer set of environmental information~\cite{jia2016enhanced}. A unified V2V and V2I hybrid control structure should be further exploited. Moreover, a well-designed hybrid structure can also help to deal with the cyber-security challenges~\cite{zhu2009security}.

%\vspace{0.05in}

%\vspace{0.1in}
\subsubsection{Communication Imperfection Resilient Control Scheme}
\vspace{0.01in}

%\noindent \textbf{Communication imperfection resilient control scheme}~

\FL{ Although the 5G technology can achieve {\acpt a} high communication bandwidth with low latency{\acpt ,} the chances of information loss are still not negligible especially when the cyber-attack becomes the new threat to the V2X communications.} It is also demonstrated that the resilient control on the longitudinal speed regulation can mitigate the imperfect communication effects and limitations efficiently ~\cite{ploeg2013graceful},~\cite{biron2018real}. How to design the resilient control scheme for other layers (i.e., routing planning layer and driving mode selection layer) is another important future research direction.  In addition, the optimal scheduling algorithms for V2X message transmissions based on the urgency of the vehicular tasks can also improve the overall performance of the CAV systems under the limited communication capability, which might be another solution to the imperfect communication effects and limitations.

%\vspace{0.1in}
\subsubsection{Driving States Prediction/Training of Non-connected Vehicles.}
\vspace{0.01in}

%\vspace{0.05in}
%\noindent \textbf{Driving states prediction/training of non-connected vehicles.}~

\FL{For the information-aware controller designs, vehicle trajectory optimization is the
core objective. In the mixed traffic scenarios, non-connected and non-automated vehicles are considered as the sources of uncertainty within the control input calculation process. If the manual
driving states can be predicted accurately within a reasonable time horizon, the performance of the trajectory optimization algorithm can be definitely improved~\cite{li2017cooperative}. Moreover, the driver assistant
system can also train/direct the human driver within a certain driving pattern~\cite{bojarski2017explaining}. Hence, the operation of the non-connected vehicles could be more predictable.}

\section{Conclusions}
%\vspace{0.1in}
\label{sec:conclusion}

A CAV can improve the roadway safety and operational efficiency significantly by reducing the human errors of manual driving. We conducted an in-depth review of three key factors (sensing and communication technologies, human factors, and information-aware controller design) related to the design and implementation of the CAV systems. Based upon the critical review, the current research status, challenges, and future research directions for designing a CAV system have been identified. Firstly, we discussed different in-vehicle sensors, overall vehicular communication architectures, and their protocols with safety requirements. In addition, we presented several future generation sensor and communication technologies to be adapted for CAV systems. However, it is very important to ensure the full utilization of heterogeneous sensing and communication technologies to improve the safety and operational efficiency as well as user comfort of the CAV systems. Secondly, different human factors such as user comfort, preferences, and reliability issues were discussed in conjunction with a user acceptable CAV design. In addition, users' behavioral adaption and situation awareness were presented to give an overview of current trends. In order to accelerate the mass adoption of CAVs, the future research should be devoted towards user preferences modeling, system failure handling, and integration of multiple assistance systems for the CAV systems. Thirdly, we presented different aspects of controller design (route-planning layer, driving mode selection layer, and driving mode execution layer) with respect to the V2X communication. To further take advantages of wireless communication, future research should consider the selection between a centralized controller and a distributed controller with mixed traffic issues (e.g., manually driven and non-connected vehicles). It is expected that this paper would motivate the design of more advanced interdisciplinary technologies that integrate those sensing and communication technologies, human factors, and information-aware controller design.

\section*{Acknowledgements}
%\label {sec:Acknowledgements}
\label {sec:Acknowledgements}
%\begin{acks}

This study is supported in part by the U.S. NSF grants CCF-1822965, OAC-1724845, ACI-1719397 and CNS-1733596, Microsoft Research Faculty Fellowship 8300751, IBM Ph.D. fellowship award 2017, and U.S. Army Contracting Command-Warren under grant No. W56HZV-14-2-0001. This study is also partially supported by the Center for Connected Multimodal Mobility ($C^{2}M^{2}$) (USDOT Tier 1 University Transportation Center) headquartered at Clemson University, Clemson, South Carolina. Any opinions, findings, and conclusions or recommendations expressed in this material are those of the authors and do not necessarily reflect the views of the Center for Connected Multimodal Mobility ($C^{2}M^{2}$) and the official policy or position of the USDOT/OST-R, or any State or other entity, and the U.S. Government assumes no liability for the contents or use thereof. The authors are grateful to the anonymous reviewers for their valuable suggestions. %The authors would like to thank Mr. Liuwang Kang for his valuable discussions and comments.
%\end{acks}
%This research was supported in part by U.S. NSF grants CCF-1822965, OAC-1724845, ACI-1719397 and CNS-1733596, Microsoft Research Faculty Fellowship 8300751, and IBM Ph.D. fellowship award 2017.

\vspace{-0.1in}

%axriv
\bibliographystyle{unsrt}
%axriv

%\bibliographystyle{IEEEtran}

%\setlength{\columnsep}{30mm}
%{\footnotesize
%\bibliography{mybibfile,platoon}
%}

{\footnotesize
\bibliography{mybibfile}
}

\end{document}